\documentclass[12pt, reqno]{article}

\usepackage{jheppub}
\usepackage{subfig}
\usepackage{amssymb}
\usepackage{amsmath}
\usepackage[usenames,dvipsnames]{xcolor}
\usepackage{epsfig}
\usepackage{dcolumn}
\usepackage{tikz}
\usetikzlibrary{shapes.geometric, arrows}
\usepackage{upgreek}
\usepackage{setspace}
\usepackage{enumitem}
\usepackage{array,multirow,bigdelim,arydshln}
\usepackage{appendix}
\usepackage{xparse}
\usepackage[utf8]{inputenc}
\hypersetup{
	colorlinks,
	urlcolor=Maroon,
	linkcolor=Maroon,
	citecolor=Maroon
}

\NewDocumentCommand{\binomial}{omm}
 {%
  \genfrac(){0pt}{}{#2}{#3}%
  \IfValueT{#1}{_{\!#1}}%
 }
\NewDocumentCommand{\eulerian}{omm}
 {%
  \genfrac<>{0pt}{}{#2}{#3}%
  \IfValueT{#1}{_{\!#1}}%
 }

\def \y {y_{\textsc{b}}}
\def \x {y_{\textsc{a}}}
\def \ma {m_{\textsc{a}}}
\def \mb {m_{\textsc{b}}}
\def \one {\rm 1\mbox{-}loop}

\usepackage{latexsym}
\usepackage{tikz}

\title{Leading Singularities and Classical Gravitational Scattering}

\author[a]{Freddy Cachazo}
\author[a,b,c]{and Alfredo Guevara}

\affiliation[a]{Perimeter Institute for Theoretical Physics, Waterloo, ON N2L 2Y5, Canada}
\affiliation[b]{Department of Physics $\&$ Astronomy, University of Waterloo, Waterloo, ON N2L 3G1, Canada}
\affiliation[c]{CECs Valdivia $\&$ Departamento de F\'isica, Universidad de Concepci\'on, Casilla 160-C, Concepci\'on, Chile}
\emailAdd{fcachazo, aguevara@pitp.ca}

\abstract{In this work we propose to use leading singularities to obtain the classical pieces of amplitudes of two massive particles whose only interaction is gravitational. Leading singularities are generalizations of unitarity cuts. At one-loop we find that leading singularities obtained by multiple discontinuities in the t-channel contain all the classical information. As the main example, we show how to obtain a compact formula for the fully relativistic classical one-loop contribution to the scattering of two particles with different masses. The non-relativistic limit of the leading singularity agrees with known results in the post-Newtonian expansion. We also compute a variety of higher loop leading singularities including some all-loop families and study some of their properties.}

\begin{document}
\maketitle
\addtocontents{toc}{\protect\setcounter{tocdepth}{1}}
\def \tr {\nonumber\\}
\def \la  {\langle}
\def \ra {\rangle}
\def\hset{\texttt{h}}
\def\gset{\texttt{g}}
\def\sset{\texttt{s}}
\def \be {\begin{equation}}
\def \ee {\end{equation}}
\def \ba {\begin{eqnarray}}
\def \ea {\end{eqnarray}}
\def \k {\kappa}
\def \h {\hbar}
\def \r {\rho}
\def \l {\lambda}
\def \be {\begin{equation}}
\def \en {\end{equation}}
\def \bes {\begin{eqnarray}}
\def \ens {\end{eqnarray}}
\def \red {\color{Maroon}}

\numberwithin{equation}{section}

\section{Introduction}
\label{Intro}

Analytic properties of the S-matrix of massless particles have been intensively studied since the introduction of twistor string theory by Witten in 2003 \cite{Witten:2003nn}. One of the earliest outcomes was the recognition of ``leading singularities'' as a fundamental set of well-defined (free of divergences and gauge invariant) quantities in any field theory \cite{Britto:2004nc,ArkaniHamed:2008gz,ArkaniHamed:2009dg,ArkaniHamed:2010kv,ArkaniHamed:2012nw}. Leading singularities are generalizations of textbook unitarity cuts. While the latter compute discontinuities across codimension one branch cuts, the former correspond to singularities of the highest possible codimension \cite{analytic}.

Motivated by the spectacular advances that led to the recent gravitational wave detections \cite{TheLIGOScientific:2016src,Yunes:2016jcc}, it is natural to ask how leading singularities can be used in computations involving the scattering of two massive particles through the exchange of gravitons.

A classic textbook exercise in quantum field theory is the derivation of the Newtonian potential from the non-relativistic limit of tree-level scattering of two massive particles via a graviton. Higher order effects defining what is known as the post-Newtonian expansion are much more complicated and were studied in the same context in \cite{Duff}, followed by \cite{Hamber:1995cq,Akhundov:1996jd,Muzinich:1995uj}. Recently, a range of sophisticated techniques has led to impressive progress in effective field theory approaches \cite{Donoghue:1996kw,Scherer:2002tk,BjerrumBohr:2002kt,Blanchet:2002av,Goldberger:2004jt,Futamase2007,Holstein:08,Porto:2016pyg,Donoghue:2017pgk}. Some of them exploit both analytic and numerical techniques in order to have control over the whole evolution of binary mergers such as the very successful Effective One Body approach \cite{Buonanno:1998gg,Buonanno:2000ef,Damour:2001tu,Damour:2000we} . More recently, several applications of on-shell techniques \cite{Bjerrum-Bohr:2014lea,Bjerrum-Bohr:2013,Holstein:2016cfx,Bjerrum-Bohr:2016hpa,Bai:2016ivl}, originally developed for gluon scattering, have also been used to study the binary problem in the post-Newtonian perturbative scheme  \cite{Neill:2013wsa,Vaidya:2014kza,Holstein:2016fxh,Burger:2017yod}.

Treating the interaction of two massive bodies as a scattering process mediated by gravitons immediately separates the computation by the topology of Feynman diagrams according to their loop order. In standard field theoretic computations, loop contributions are usually related to quantum effects. In applications where at least one of the external particles is massive loop integrals can give rise to both classical and quantum effects \cite{Holstein:2004dn,Donoghue:1996mt}. There are integration regions \cite{Beneke:1997zp,Kniehl:2002br}  that can only contribute quantum effects while others can contribute to both kind of effects \cite{Neill:2013wsa}. Computations of classical potentials or effective actions thus require the separation of both effects.

In this work we explore general leading singularities in the scattering of massive particles via massless particles. Leading singularities, which are computed as multidimensional residues, generically have support outside the physical region of integration \cite{Britto:2004nc,ArkaniHamed:2008gz}. Therefore they are not naturally located on any of the regions mentioned above. However, here we argue that leading singularities associated to multiple discontinuities exclusively in the t-channel contain all the information needed to reproduce the classical scattering. Moreover, the leading singularity itself is directly computing the classical contribution as we show in  several examples.

The main example in this work are the two leading singularities that determine the full classical part of the one-loop scattering of two massive scalars with masses $\ma$ and $\mb$ exchanging gravitons. The complete fully relativistic result for one of them is expressed compactly as a contour integral
\be
\label{compactLS}
32\pi^2 G^2 \frac{\mb}{\sqrt{-t}}\frac{M^4}{\left(4-\frac{t}{\mb^2}\right)^{\frac{5}{2}}}\!\oint_{\Gamma} \!\frac{dz}{z^3}\frac{\left(\frac{1}{\sqrt{-x}}+\frac{(u-s)}{M^2}z+\sqrt{-x}z^2\right)^4}{\left(1-\frac{(\ma^2-\mb^2-s)}{M^2}
\frac{\sqrt{-t}}{\mb}z-z^2\right)\!
\left(1+\frac{(\ma^2-\mb^2-u)}{M^2}\frac{\sqrt{-t}}{\mb}z-z^2\right)}.
\ee
The contour $\Gamma$ computes the residue at $z=0$ minus that at $z=\infty$ while $M$ and $x$ are defined via the equations
\be
M^4 = (\ma^2-\mb^2)^2-s u \quad {\rm and} \quad (1+x)^2=\frac{t}{\mb^2}x.
\ee
Here $s,t,u$ are standard Mandelstam invariants satisfying $s+t+u=2(\ma^2+\mb^2)$.

The leading singularity computed by \eqref{compactLS} has the topology of a triangle with a massive $\mb$ propagator (see fig. \ref{fig:gravtrian} in section 2).

The second leading singularity is the one corresponding to its reflection and it is obtained by simply exchanging $\ma$ and $\mb$.

The non-relativistic limit of \eqref{compactLS} in the center of mass frame is
\be
16G^2\pi^2  \frac{\mb}{|\vec{q}|}\left( 6 \ma^2\mb^2 +\frac{15}{2}(\ma+\mb)^2\vec{p}^{\,2}+{\cal O}\left(|\vec{p}|^4\right) \right) + {\cal O}\left(|\vec{q}|^0\right),
\ee
where $\vec{q}$ is the momentum transfer, while $\vec{p}$ is the average momentum of the system. 

In order to make the link between classical pieces of scattering amplitudes and their leading singularities, we propose a construction using multiple dispersion relations in the t-channel which projects out irrelevant information. In fact, \eqref{compactLS} is nothing but the double discontinuity across branch cuts in the t-channel of the one-loop amplitude. It turns out that the leading singularity remains invariant, up to terms projected out by the construction, after being integrated along the branch cuts in the dispersion integrals. Single dispersion integrals have been used as a tool in the computation of corrections to classical potentials for a long time (see e.g. the work of Feinberg and Sucher \cite{Feinberg:1988yw}) but we find that multiple dispersion integrals provide a natural way of separating classical from quantum contributions.

The final step is to add \eqref{compactLS} to the contribution from the reflected leading singularity and include a non-relativistic normalization to obtain\footnote{A factor of $4$ has been included which comes from the dispersion relations. This is explained in detail in section 3.}
\be\label{potentialOneLoopintro}
\frac{M^{\one}_{\rm classical}(t)}{4E_\textsc{A}E_\textsc{B}} = G^2\pi^2  \frac{(\ma+\mb)}{|\vec{q}|}\left(6\ma \mb +\frac{9(\ma^2+\mb^2)+30 \ma \mb}{2\ma \mb}\vec{p}^{\,2} +{\cal O}\left(|\vec{p}|^4\right)\right).
\ee
This is the known one-loop contribution to the classical part of the normalized amplitude \cite{Neill:2013wsa}. From this formula there is a standard procedure to obtain the classical potential, $V^{\rm 1-loop}_{\rm classical}$, in the post-Newtonian expansion\footnote{The reason \eqref{potentialOneLoopintro} is not $V^{\rm 1-loop}_{\rm classical}$ is that to this order in coupling constants one has to subtract the contribution from the iteration of the tree-level potential.} (see e.g. \cite{Neill:2013wsa} for details).

This paper is organized as follows. The main purpose of this work is to introduce the concept of leading singularities in the context of gravitational scattering so in section 2 the general definition of leading singularities is introduced and illustrated via a variety of examples. We start with theories containing only scalar particles where computations are simpler and then move on to gravitational scattering. Conveniently, computations in scalar theories provide useful intermediate results for their gravitation counterparts. All examples are at one-loop with two- and higher loop cases postponed until section 4. In section 3, we concentrate on the problem of reproducing the classical contributions to the scattering of two massive particles via gravitons using the results from section 2. In order to set the stage, we start with the tree-level computation using a BCFW recursion relation construction which is then linked to a dispersion relation in the t-channel. This is used to motivate a double dispersion projection in the t-channel at one-loop which leads to the connections presented above between \eqref{compactLS} and the classical contribution \eqref{potentialOneLoopintro}. In section 4, we provide several results on leading singularities. These include a more formal connection between leading singularities and multiple discontinuities which justifies their use in section 3 as well as examples of two and higher loop leading singularities. In section 5, we end with discussion which include some possible future directions.

\section{Leading Singularities in General Theories}

Scattering amplitudes possess a very intricate analytic structure in perturbation theory as can be seen from imposing unitarity \cite{analytic}. When the unitarity constraint is imposed in a given channel, it relates the discontinuity of the amplitude to the exchange of on-shell states between two sets of external particles. The one-particle exchange implies the existence of poles while a two-particle exchange implies the presence of a branch cut. In most cases, the discontinuities in a given channel also possess an intricate analytic structure and the process can be repeated leading to what is known as generalized unitarity constraints \cite{analytic}.

The discontinuity across a pole is simply the residue at the location of the pole. Most quantum field theory textbooks present discontinuities in a given channel from two-particle exchanges and refer to them as unitarity cuts. These can also be thought of as residues of the amplitude by taking two propagators $1/(L_1^2-m_1^2+i\epsilon)$ and $1/(L_2^2-m_2^2+i\epsilon)$ to define variables $1/u_1$ and $1/u_2$ and integrate over contours $|u_a|=\varepsilon$ that encircle $u_a=0$ in the corresponding complex planes.
This process is usually known as ``cutting'' propagators\footnote{The cutting process also involves a step function $\theta(L^0_a)$ but this will not play a role in this section. These step function will be crucial in section 4 and we postpone their introduction until then.}. The term comes from the fact that this is equivalent to removing the principal part of $1/(L_a^2-m_a^2+i\epsilon)$ while keeping the delta function imposing the on-shell condition $L_a^2=m_a^2$.

Generalized unitarity explores further discontinuities and these too can be realized as contour integrals. Every time a residue is computed one explores a higher codimension singularity. The maximal number of residues at $L$-loop order in four dimensions is $4L$. Taking $4L$ residues gives rise to the highest codimension singularity and its discontinuity is known as the {\it leading singularity} \cite{Britto:2004nc,ArkaniHamed:2008gz,ArkaniHamed:2009dg,ArkaniHamed:2010kv,ArkaniHamed:2012nw}.

\begin{figure}
\centering
    \includegraphics[scale=0.50]{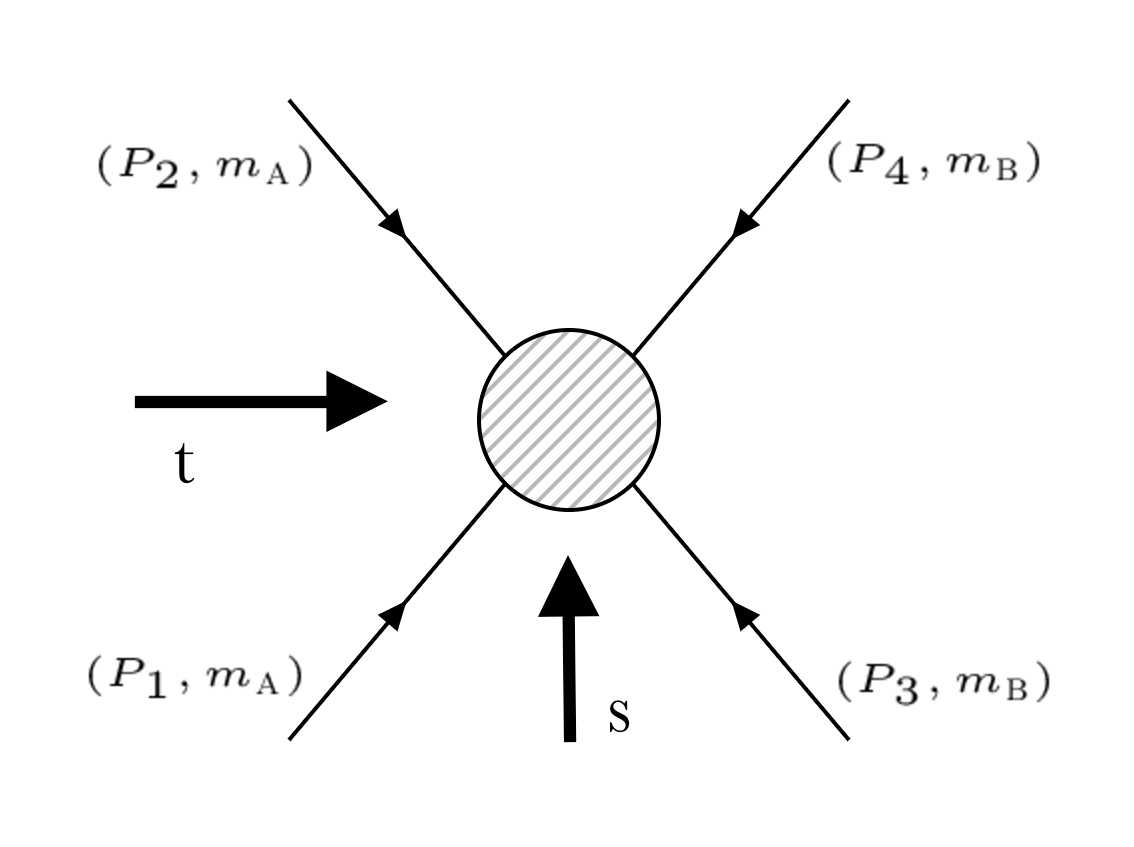}
\caption{Scattering process of massive scalars. Throughout the text we will extensively discuss the case $m_1=m_2=\ma$ and $m_3=m_4=\mb$. Note that all momenta are incoming. }\label{blob}
\end{figure}


Standard unitarity cuts can have divergences and might need a regulator. Divergences come from integrals performed over non-compact contours. Leading singularities are computed using only compact contours and are therefore finite. Also, just as unitarity cuts, leading singularities only involve physical states and are gauge invariant (see e.g. \cite{ArkaniHamed:2012nw}). These features make them ideal quantities to study in general theories.

Before proceeding to the computation of leading singularities in gravitational scattering, we start with leading singularities in a theory with a massive and a massless scalar field. Keeping in mind the applications to gravitational scattering we restrict the study to amplitudes with four external states. The scattering picture for massive particles is represented in fig. \ref{blob}, where our conventions are set to all-incoming states. Hereafter we denote by $k_i$ the momenta associated to massless particles, while $P_i$ will denote the external momenta for massive ones.

\subsection{Leading Singularities in Scalar Theories}
\label{scalar}

In this section we consider a variety of scalar theory leading singularities. Some scalars have a mass while others are massless. Interaction terms are taken to be of all possible orders, i.e., cubic, quartic, etc. The reason is that we are interested in the most general leading singularities that can be present in gravitational interactions.

\begin{figure}
\centering
    \subfloat[]{{\includegraphics[width=7cm]{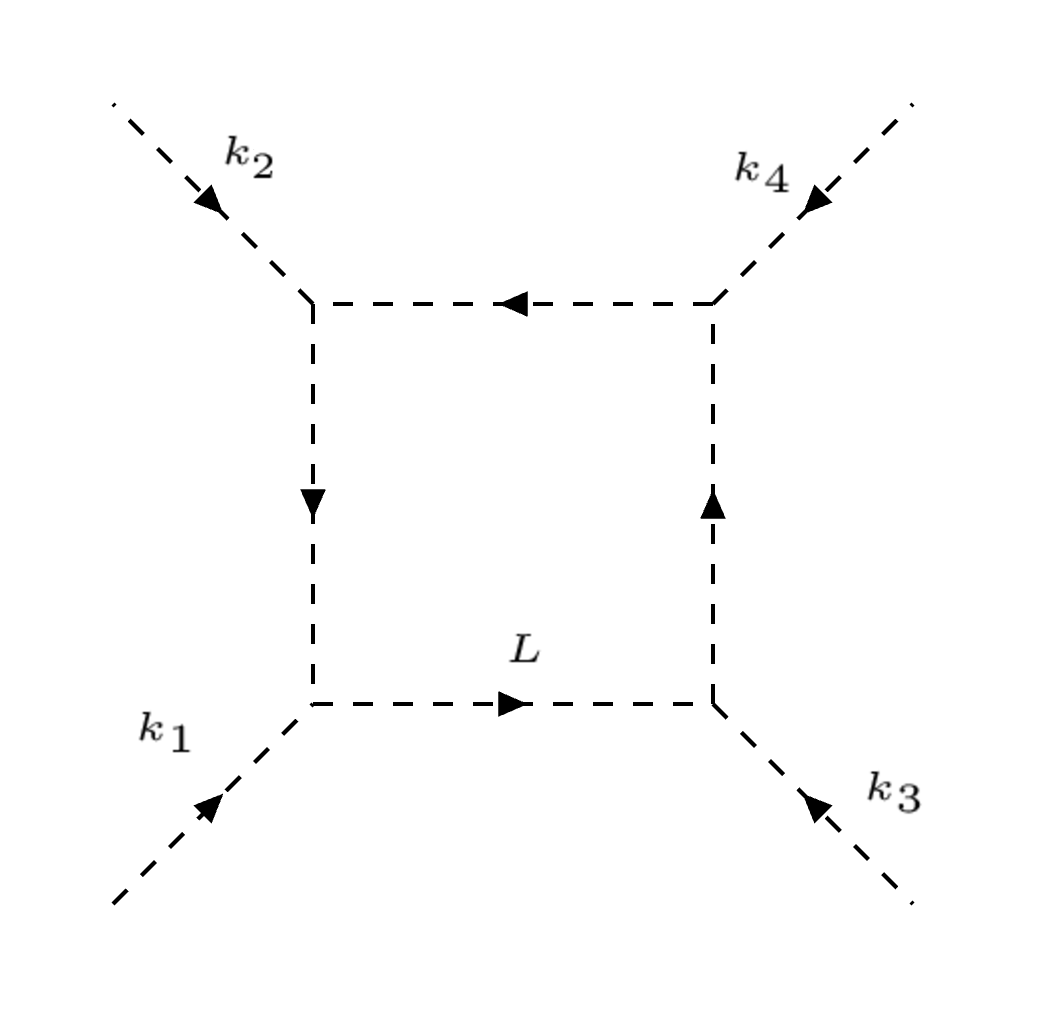} }}%
    \qquad
    \subfloat[]{{\includegraphics[width=7cm]{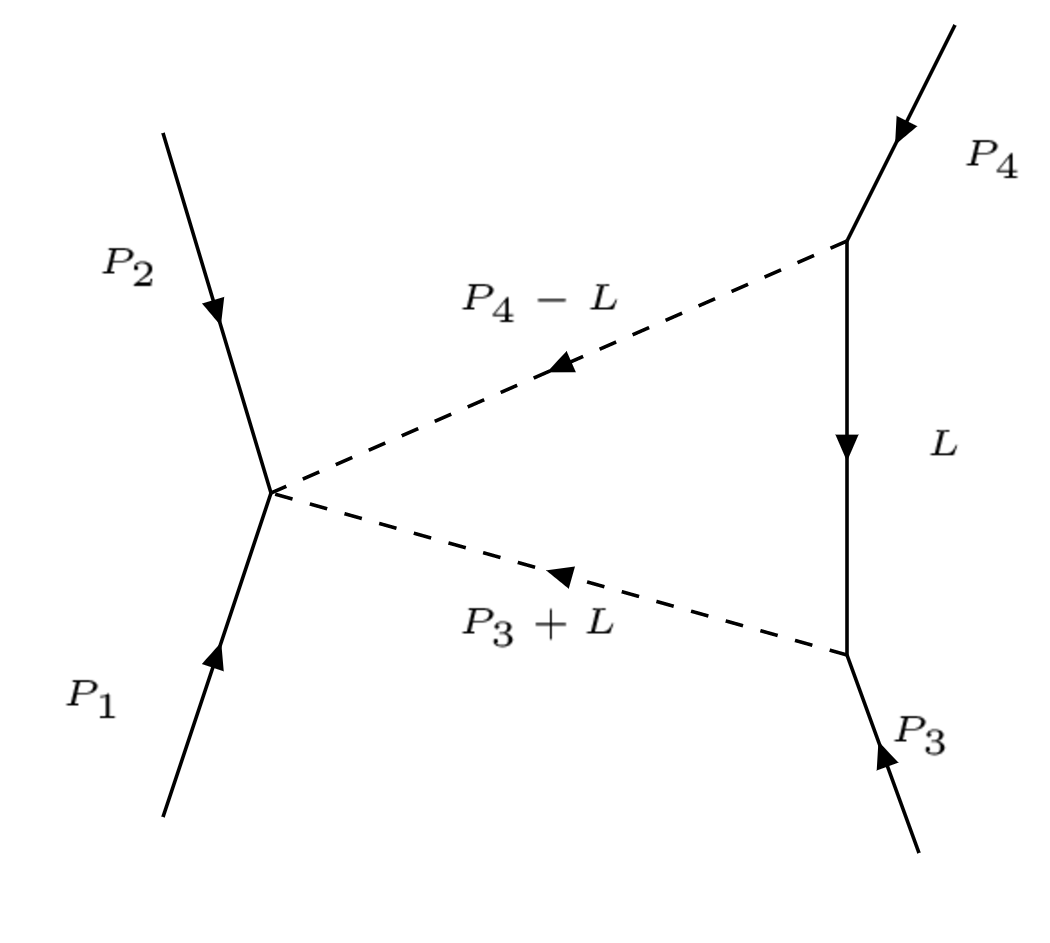} }}%
    \caption{a) Box with internal and external massless particles. b) Triangle with massive external particles and massless internal particles. All lines in this work represent on-shell particles.}%
    \label{fig:triangle1}
\end{figure}

\paragraph{Massless box diagram.}
One of the simplest examples is the leading singularity of a one-loop four particle amplitude in a massless scalar theory with trivalent interactions, see fig. \ref{fig:triangle1}a. The leading singularity is given by a contour integral of the form
\be
{\rm LS}=\int_{\Gamma_{\rm LS}}d^4L \frac{M_3(L_1,k_1,L_2)M_3(L_2,k_2,L_3)M_3(L_3,k_4,L_3)M_4(L_4,k_3,L_1)}{L^2(L-k_1)^2(L+k_3)^2(L-(k_1+k_2))^2}
\ee
where $L_1=L$, $L_2=L-k_1$, $L_3=L-(k_1+k_2)$ and $L_4=L+k_3$. The contour $\Gamma_{\rm LS}$ has the topology of $(S^1)^4$ and it is defined by $|L^2|=\epsilon$, $|(L-k_1)^2|=\epsilon$, $|(L+k_3)^2|=\epsilon$, and $|(L-(k_1+k_2))^2|=\epsilon$. Each $M_3(p,q,r)$ is a fully on-shell three-particle tree amplitude of the theory. In this example $M_3(p,q,r)$ is simply given by the cubic coupling constant of the theory $g_3$.

There are two contours $\Gamma_{\rm LS}$. Using the spinor helicity formalism\footnote{Any four-vector $P^{\mu}$ can be transformed into a bispinor by using the four-vector of Pauli matrices $P^\mu \sigma^\nu_{\alpha\dot\alpha}\eta_{\mu\nu}$. Given two spinors $\lambda_\alpha$ and $\lambda'_\alpha$, the $SL(2,\mathbb{C})$ invariant product is denoted by $\langle\lambda~\lambda'\rangle := \lambda_\alpha\lambda'_{\beta}\epsilon^{\alpha\beta}$. Likewise for spinors of the opposite helicity one has $[\tilde\lambda~\tilde\lambda']$.}, all external massless particle momenta can be written as $(k_a)_{\alpha,\dot\alpha}=(\lambda_a)_\alpha(\tilde\lambda_a)_{\dot\alpha}$. The loop momentum at the location of the poles becomes either $L_{\alpha,\dot\alpha} = \frac{[1\; 2]}{[3\; 2]}(\lambda_1)_\alpha(\tilde\lambda_3)_{\dot\alpha}$ or $L_{\alpha,\dot\alpha} =\frac{\langle 1\; 2\rangle}{\langle 3\; 2\rangle} (\lambda_3)_\alpha(\tilde\lambda_1)_{\dot\alpha}$. Note that on the physical contour $\mathbb{R}^4$, this integral is IR divergent and needs a regulator.

The contour integral is easily performed and gives $g_3^4\times 1/st$ with $s=(k_1+k_3)^2$ and $t=(k_1+k_2)^2$.

The final comment on this example is that the result $1/st$ is what it would have been obtained by computing the integral, using e.g. dimensional regularization, and then evaluating the discontinuity across the t-channel branch cut and the discontinuity of the result across the s-channel branch cut \cite{Britto:2004nc}. Building on the intuition from standard unitarity cuts, the t-channel discontinuity is computed by cutting $1/L^2$ and $(L-(k_1+k_2))^2$ while that in the s-channel by cutting $1/(L-k_1)^2$ and $1/(L-(k_1+k_2))^2$. Here we have performed both simultaneously.

\paragraph{Triangle with massive external particles.}
The next example is the one-loop scattering of four massive scalars interacting via the exchange of massless scalars $\phi$. Let us assume that there are two kind of massive scalars fields $\Phi_A$ and $\Phi_B$ with masses $\ma$ and $\mb$. They each have only one kind of interaction vertex with the massless scalars. We take them to be quartic and cubic couplings respectively (see fig. \ref{fig:triangle1}). The leading singularity is computed by the integral
\be
{\rm LS}=\int_{\Gamma_{\rm LS}}d^4L \frac{M_4(L_1,P_1,P_2,L_2)M_3(L_2,P_4,L_3)M_3(L_3,P_3,L_1)}{(L+P_3)^2(L^2-\mb^2)(L-P_4)^2}
\ee
where $L_1=L$, $L_2=P_4-L$ and $L_3=L+P_3$. Once again, the on-shell amplitudes are all given by the coupling constants $M_4(\phi,\Phi_A,\Phi_A,\phi)=g_4$, $M_3(\phi,\Phi_B,\phi)=g_3$.

It might be surprising that there is a leading singularity contour with four poles while the integral only has three propagators. As it will be clear from the computation, a new pole, not visible at first, appears when some propagators are cut \cite{Buchbinder:2005wp}.

The contour integral becomes
\be
I=\int_{\Gamma_{\rm LS}} \frac{d^4L}{(L^2-\mb^2)(L+P_3)^2(L-P_4)^2}.
\ee
Here $P_3^2=P_4^2=\mb^2$. Note that the loop momenta $L$  is associated to the massive propagator. In order to compute the integral, and for future convenience, we introduce a suitable parametrization of $L$
\be
L= z \ell + \omega q\,,\quad \ell_{\alpha,\dot{\alpha}}=\lambda_{\alpha}  \tilde{\lambda}_{\dot{\alpha}}\,.
\ee
Here the integration variables correspond to the scales $z, \omega \in \mathbb{C}$ and the (projective) spinors $\lambda_{\alpha}$,  $\tilde{\lambda}_{\dot{\alpha}}$. All together they parameterize $L\in \mathbb{C}$ while $q$ is a fixed reference massless vector. Cutting $L^2-\mb^2$ means that we are dealing with the Lorentz invariant phase space integral of a massive vector. It is well-known that the measure becomes \cite{Nair:1988bq, Brandhuber:2004yw, Cachazo:2004kj, Britto:2004nc}
\be
\frac{1}{(2\pi)^4}\frac{d^4L}{(L^2-\mb^2)} = \frac{1}{(2\pi)^4} z\; dz\; \langle \lambda~ d\lambda\rangle [\tilde\lambda~ d\tilde\lambda]\,\frac{d\omega}{4(\omega- \frac{\mb^2}{2z\ell\cdot q})}.
\ee
 where we expanded the massive propagator using $L^2= 2z\omega\ell\! \cdot\! q$. For convenience we have restored the factor $(2\pi)^4$ from the measure and used $\langle \lambda~ d\lambda\rangle$ to denote $\epsilon^{\alpha\beta}\lambda_\alpha d\lambda_\beta$. The integral around the pole $L^2-\mb^2=0$ can then trivially done as a contour integral in the $\omega$ plane, extracting the corresponding residue. This fixes $\omega=\frac{\mb^2}{2z\ell\cdot q}$, while the leading singularity takes the form
 \be
I=\frac{1}{4(2\pi i )^3}\int_{\Gamma_{\rm LS}} \frac{z dz \langle \lambda~ d\lambda\rangle [\tilde\lambda~ d\tilde\lambda]}{(2\mb^2+ 2L\cdot P_3)(2\mb^2- 2L\cdot P_4)}\,.
\ee
In order to compute the residue around the massless propagators we introduce at this stage two auxiliary massless vectors, $(p_3)_{\alpha,\dot{\alpha}}=(\lambda_3)_{\alpha}  (\tilde{\lambda}_3)_{\dot{\alpha}}$ and $(p_4)_{\alpha,\dot{\alpha}}=(\lambda_4)_{\alpha}  (\tilde{\lambda}_4)_{\dot{\alpha}}$,  which satisfy the relations
\begin{eqnarray}
\label{little}
P_3= p_3 + x p_4\,, \quad
P_4=p_4+ x p_3\,, \quad x=\dfrac{\mb^2}{2p_3 \cdot p_4} \,.
\end{eqnarray}
The last equation is just the on-shell condition for $P_3$ and $P_4$. It is easy to verify that
\be
\label{xandt}
\dfrac{(1+x)^2}{x} = \dfrac{t}{\mb^2}\quad \mbox{or equivalently}\quad \dfrac{(1-x)^2}{x} = \dfrac{t-4\mb^2}{\mb^2}\, ,
\ee
so that $x$ can be regarded as a useful parametrization of the $t$-channel as hinted already in the introduction\footnote{The transformation \eqref{little} is invertible except at the singular points $x^2=1$, corresponding to the physical threshold $t=0,4\mb^2$. We will come back at this for the discussion of gravitational scattering.}. Now we can choose the reference vector to be $q_{\alpha,\dot{\alpha}}=(\lambda_3)_{\alpha}  (\tilde{\lambda}_4)_{\dot{\alpha}}$. We also define its conjugate $\bar{q}_{\alpha,\dot{\alpha}}=(\lambda_4)_{\alpha}  (\tilde{\lambda}_3)_{\dot{\alpha}}$. As these are linearly independent, we expand
\be
\ell= A p_3 + B p_4+ C q + D \bar{q}\,.
\ee
The overall scale of $\ell$ is irrelevant as it can be absorbed into $z$ and it can be used to set $D=1$. Imposing $\ell^2=0$ fixes $C=AB$. Now we regard $A, B\in \mathbb{C}$ as the integration variables corresponding to the measure $\langle \lambda ~d\lambda\rangle [\tilde\lambda ~d\tilde\lambda]$. Performing the change of variables leads to
\be
I=\frac{1}{(2\pi i)^3} \dfrac{(2p_3 \cdot p_4)}{16} \int_{\Gamma_{\rm LS}}\dfrac{zdz\, dA\, dB}{(\mb^2+zp_3\cdot p_4(B+xA))(-\mb^2+zp_3\cdot p_4 (A+xB))}\,.
\ee
The location of the poles for the two propagators corresponds to $A=-B=\frac{2x}{z(1-x)}$. Performing the integrals finally leaves
\begin{align}
\label{measure}
I=& \dfrac{x}{4\mb^2(1-x^2)}\left(\frac{1}{2\pi i }\int_{\Gamma_{\rm LS}} \dfrac{dz}{z}\right)\,.
\end{align}
We note the presence of emergent poles at both $z=\infty$ and $z=0$, directly arising from the integration measure of the triple-cut. In this case both poles yield the same contribution for the leading singularity. Thus, we are now in a position to define $\Gamma_{\rm LS}$ as the contour enclosing either of these, and the leading singularity as the corresponding residue. In section \ref{discont} we show how these poles arise naturally in a completely different parametrization, and we will discuss the meaning of the leading singularity as a second discontinuity operation in the $t$-channel.\\

By choosing the contour at $z=\infty$, i.e. $\Gamma_{\rm (S^1)^3\times S^1_\infty}$, and using the definition of $x$ we can write the final result as
\be\label{trian}
{\rm LS}_{\rm triangle}= g_4 (g_3)^2 \int_{\Gamma_{\rm (S^1)^3\times S^1_\infty}} \frac{d^4L}{(L^2-\mb^2)(L+P_3)^2(L-P_4)^2} = \frac{g_4 (g_3)^2}{4\sqrt{(-t)(4\mb^2-t)}}.
\ee
This leading singularity can be iterated in order to compute one with an arbitrary number loops where scalar triangles are arranged in a nested topology (fig.\ref{double}). In section 4 we present more details on this and other higher loop examples.

\paragraph{Box with massive external particles.}
For our final example in this section consider the case in which both massive scalars can only interact with the massless scalars via three particle couplings. At one-loop, the amplitude with four external massive scalars gives rise to the following contour integral (we suppress the three-particle amplitudes as they are all given by the coupling constant),
\be
{\rm LS}_{\rm box}=\int d^4L \frac{1}{(L+P_3)^2((L+P_3+P_1)^2-\ma^2)(L^2-\mb^2)(P_4-L)^2}.
\ee
This box contour integral is easy to compute using the previous parametrization. In fact, cutting the three propagators associated to the triangle leads to the measure \eqref{measure}. We now only need to include the fourth propagator, which we write in terms of the new variables
\begin{align}
{\rm LS}_{\rm box}=& \dfrac{x}{4(1-x^2)\ma^2}\left(\frac{1}{2\pi i }\int_{\Gamma_{\rm LS}} \dfrac{dz}{z}\dfrac{1}{(L+P_3+P_1)^2-\ma^2}\right)\, \nonumber \\
=&\dfrac{x}{8(1+x)^2 \mb^2}\left(\frac{1}{2\pi i }\int_{\Gamma_{\rm LS}} \dfrac{dz}{z^2(\frac{1-x}{1+x})\bar{q}\cdot P_1+ z (p_3-xp_4)\cdot P_1-x(\frac{1+x}{1-x})q\cdot P_1}\right)\, \nonumber \\
=&\dfrac{1}{8t}\left(\frac{1}{2\pi i }\int_{\Gamma_{\rm LS}} \dfrac{dy}{y^2(q\cdot P_1\bar{q}\cdot P_1)+ y (p_3-xp_4)\cdot P_1-x}\right)\,,
\end{align}
where a change of variables $z=y(\frac{1+x}{1-x})$ was used in the last equality. Note that the poles at $z=0$ and $z=\infty$ associated to the triangle leading singularity are replaced by two poles associated to the additional massive propagator. Also the non-analytic (i.e. containing a $\sqrt{-t}$) prefactor is replaced by $\frac{x}{(1+x)^2 \mb^2}=\frac{1}{t}$. Again, there are two possible contours defined by
\be
|y^2(q\cdot P_1\bar{q}\cdot P_1)+ y (p_3-x p_4)\cdot P_1-x| = \epsilon
\ee
and centered around the two roots of the quadratic polynomial. Here we define the leading singularity as the residue at one of the two poles. Clearly, the residue at the other pole only differs by a sign as the integral has no other poles and the sum over the two residues must vanish.

For future convenience, let us define the quantity $M$ as a solution to the equation
\be
\label{M}
M^4 := -4(1-x)^2(q\cdot P_1\bar{q}\cdot P_1) = (\ma^2-\mb^2)^2-s u \,,
\ee
where we have used momentum conservation, $s+t+u=2(\ma^2+\mb^2)$ together with \eqref{little} (see also eq. \eqref{p3-xp4} below).
Restoring the corresponding couplings, we find the leading singularity to take the form
\begin{align}
\label{scalarboxls}
 {\rm LS}_{\rm box}=&\frac{(g_3)^4}{\sqrt{M^4 - st}}\times \frac{1}{t} \, \nonumber \\
=&\frac{(g_3)^4}{\sqrt{(s-\ma^2-\mb^2)^2-4\ma^2\mb^2}}\times \frac{1}{t}\,.
\end{align}
This computation is also the basis for an infinite family of leading singularities with the topology of a ladder with $r+1$ rungs, see fig. \ref{rbox}. We discuss this case in section 4 where it is shown how one can find $4r$ poles in an integral that only possesses $3r+1$ propagators following the construction introduced in \cite{Buchbinder:2005wp}.

\subsection{Leading Singularities in Gravitational Scattering}
\label{sec:gravscat}

In this section we compute several examples of leading singularities in the scattering of massive scalars interacting via gravitons. As shown in the scalar examples, leading singularities are built using on-shell amplitudes. Let us list all the relevant tree amplitudes that will be used in this section. Scalar particles $\Phi_A$ and $\Phi_B$ again have masses $\ma$ and $\mb$ while gravitons of positive and negative helicity are denoted by $G^{\pm}$. Denoting the momenta of the gravitons by $k_i$ and  introducing $\kappa= \sqrt{32\pi G}$, we have
\begin{align}
M_3(\Phi_1,\Phi_2,G^+) =\frac{\kappa}{2}    \dfrac{\langle q |P_1|k]^2  }{ \langle q~k \rangle ^2}    ,&\quad M_3(\Phi_1,\Phi_2,G^-) =\frac{\kappa}{2}     \dfrac{[q|P_1|k\rangle^2  }{ [q~k]^2} ,\quad \label{treeamp} \\
M_4(\Phi_1,\Phi_2,G_3^+,G_4^-) =& \frac{\kappa^2}{4}    \dfrac{1}{(P_1 + P_2)^2 }\dfrac{[k_3|P_1|k_4\rangle ^4}{[k_3|P_1|k_3\rangle[k_3|P_2|k_3\rangle} \,,  \label{4pt} \\
M_4(\Phi_1,\Phi_2,G_{3}^{+},G_{4}^{+})
=&     \dfrac{\kappa^2}{4} \dfrac{\ma^4}{(P_1 + P_2)^2 }\dfrac{[k_3~ k_4] ^4}{[k_3|P_1|k_3\rangle[k_3|P_2|k_3\rangle} \, . \label{m4sameh}
\end{align}
The motivation for the notation $\langle q |P|k] := (\lambda_q)_\alpha P^{\alpha\dot\alpha}(\tilde\lambda_{\dot\alpha})$ can easily be explained by noting that when $P$ is replaced by a null vector $p$ then $\langle q |p|k]=\langle q\;p\rangle [p\;k]$. These amplitudes have been computed in a variety of ways in the literature (see e.g. \cite{Neill:2013wsa,Burger:2017yod}) and they require the introduction of a reference null vector $q$. It is easy to show that the amplitudes are independent of the choice up to momentum conservation in exactly the same way as they would be gauge invariant when written in terms of polarization vectors.

\begin{figure}
  \centering
    \includegraphics[scale=0.45]{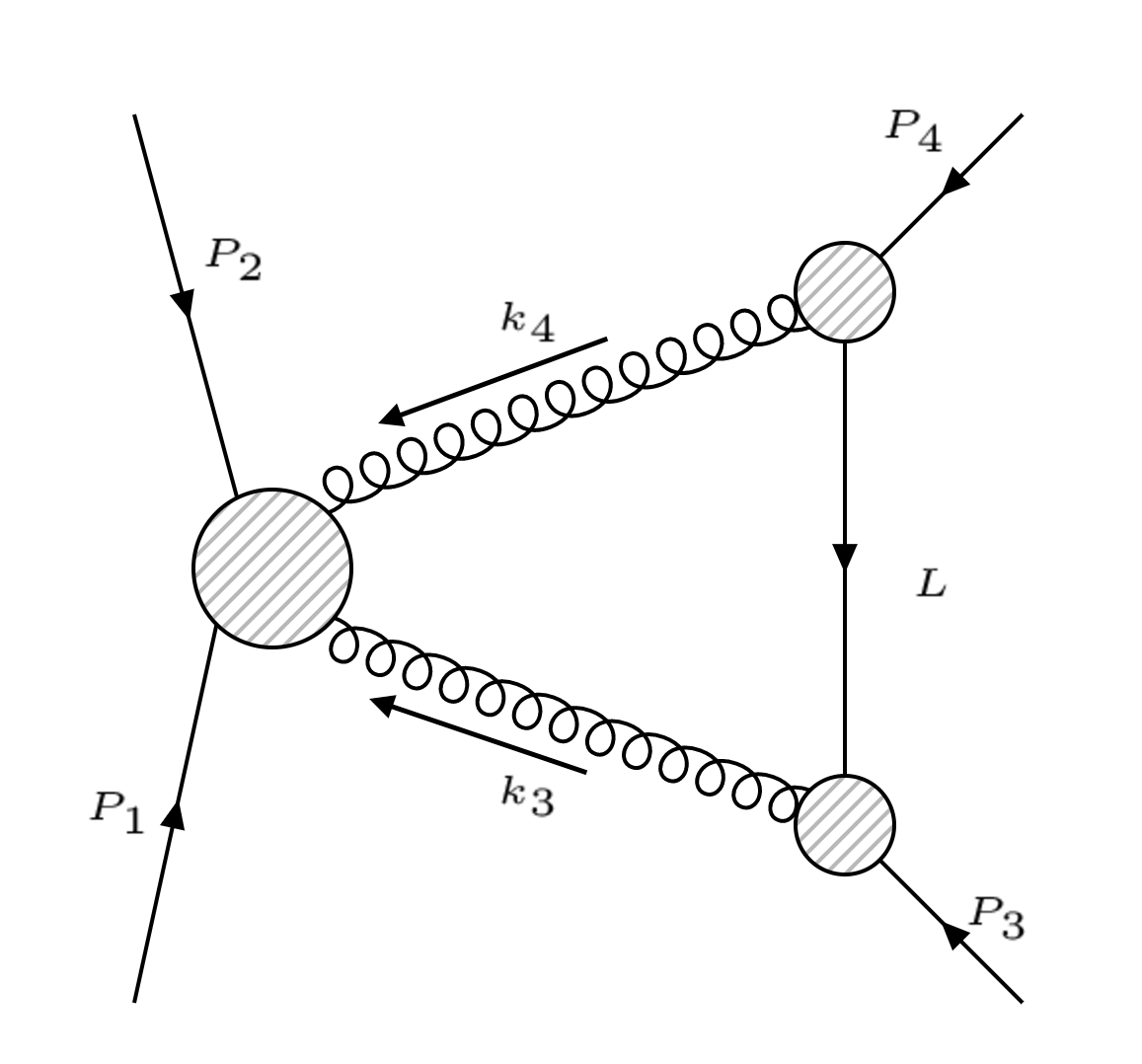}
    \caption{The box and triangle leading singularities can be computed from the triple-cut diagram.}\label{fig:gravtrian}
\end{figure}

In this section we compute two one-loop leading singularities. The first is the analog of the triangle topology in the purely scalar case while the second is the box topology. In the previous section the corresponding contours were defined. The main difference here is that unlike the purely scalar case, all tree-amplitudes are non-trivial and therefore modify the computation in interesting ways. The starting point for both computations is the same and it is given by the contour integral
\be
\label{gravitytriplecut}
I\!=\!\!\sum_{h_3,h_4} \int\!\!   \frac{d^4L}{L^2-\mb^2}\frac{1}{(L+P_3)^2(L-P_4)^2}M_4(P_1,P_2,k_3^{h_3}\!,k_4^{h_4})M_3(k_4^{-h_4}\! ,P_4,L)
M_3(L,P_3,k_3^{-h_3}\!)
\ee
where $k_3 = L+P_3$ and $k_4 = L-P_4$. The sum is over all possible helicity configurations (see figure \ref{fig:gravtrian}). Such a sum decomposes the leading singularity into four pieces, $I_{h_3,h_4}$, as follows
\be
I = I_{-+} + I_{+-} +I_{++}+ I_{--}.
\ee

We start with $h_3=-2, h_4=+2$. The configuration $h_3=-h_4=+2$ is then obtained by conjugation. At the end we briefly explain how the $h_3=h_4$ configurations have zero contribution to the triangle leading singularity.

Performing now the triple cut in the visible propagators we again obtain the measure \eqref{measure}, leading to
\be
\label{gravtrian}
I_{-+}=\dfrac{x}{4(1-x^2)\mb^2}\left(\frac{1}{2\pi i }\int_{\Gamma_{\rm LS}} \dfrac{dz}{z}M_4(P_1,P_2,k_3^{-},k_4^{+})M_3(k_4^{-},P_4,L)M_3(L,P_3,k_3^{+})\right)\,.
\ee
In order to compute the $z$ integral using \eqref{treeamp} we need to provide expressions for the momenta $k_i$, $i=3,4$, and their corresponding spinor variables.  Note that any little group transformation on these cancels in \eqref{gravtrian}, as they correspond to internal particles. Thus we can freely choose the spinor variables by arbitrarily decomposing the momenta $k_3$ and $k_4$. These can be readily computed using the parametrization of section \ref{scalar}, giving
\begin{align}
k_3(z)=&\underbrace{r(x)\left(\lambda_3 + \frac{z}{r(x)}\lambda_4 \right)}_{\lambda_{k_3}}\underbrace{\left(\tilde{\lambda}_3-\frac{x}{z}r(x)\tilde{\lambda}_4 \right)}_{\tilde{\lambda}_{k_3}}\,, \nonumber
\end{align}
\begin{align}
\label{k3k4}
k_4(z)=&\underbrace{r(x)\left( \lambda_4 + \frac{x}{z}r(x)\lambda_3 \right)}_{\lambda_{k_4}}\underbrace{\left(\tilde{\lambda}_4-\frac{z}{r(x)}\tilde{\lambda}_3 \right)}_{\tilde{\lambda}_{k_4}}\,,
\end{align}
where $r(x)=(1+x)/(1-x)$. Here we have suppressed the spinor indices since all quantities involved are $2\times 2$ matrices. We can now compute
\begin{align}
M_3(k_4^{-},P_4,L)M_3(L,P_3,k_3^{+}) =& \dfrac{\kappa^2}{4}  \left( \dfrac{\langle \bar{q} |P_3|k_3]^2  }{ \langle \bar{q}~k_3 \rangle ^2} \right) \left(  \dfrac{[\bar{q}|P_4|k_4\rangle^2  }{ [\bar{q}~k_4]^2}  \right) \, \nonumber \\
=& \dfrac{\kappa^2}{4} \left ( 2p_3 \cdot p_4 \dfrac{\langle k_4 ~p_4 \rangle [k_3 ~p_3]}{[k_4~ p_3]\langle p_4 ~k_3\rangle }\right) ^2 \, \nonumber \\
=& \dfrac{\kappa^2}{4} \dfrac{\mb^4 x^2}{z^4}r(x)^4\,.
\end{align}
In the three-point amplitudes we have chosen the reference spinors corresponding to the vector $\bar{q}$. Plugging this into \eqref{gravtrian} we find our first main result
\be
\label{mresult}
I_{-+}=\dfrac{\kappa ^2 \mb^2x^3}{16(1-x^2)}\left(\dfrac{1+x}{1-x}\right)^4 \left(\frac{1}{2\pi i }\int_{\Gamma_{\rm LS}} \dfrac{dz}{z^5}M_4(P_1,P_2,k_3^{-},k_4^{+}) \right)\,.
\ee
At this stage note that this formula is very general as nothing relating to the identity of the particles $\Phi_1$ and $\Phi_2$ has been used. This means that one can choose $M_4(P_1,P_2,k_3^{-},k_4^{+})$ according to the problem in consideration. In principle one can replace the scalar particles by any two particles with given mass (including massless) and spin and compute the respective leading singularity. In this work we are interested in massive scalars undergoing gravitational scattering. This means that $M_4(P_1,P_2,k_3^{-},k_4^{+})$ is the amplitude given in \eqref{4pt} for a scalar particle of mass $\ma$. In section 4 we will also use this expression in order to explore the leading singularity associated to a two-loop diagram.

Returning to the computation,  using \eqref{k3k4} we have
\begin{align}
M_4(P_1,P_2,k_3^{-},k_4^{+})=& \dfrac{\kappa^2}{4} \dfrac{1}{(P_1 + P_2)^2 }\dfrac{[k_3|P_1|k_4\rangle ^4}{[k_3|P_1|k_3\rangle[k_3|P_2|k_3\rangle} \,  \nonumber \\
=& \kappa^2 \dfrac{z^2}{(P_1 + P_2)^2}\left(\dfrac{1+x}{1-x}\right)^2 \times  \nonumber \\
&\dfrac{ z^2(\frac{1-x}{1+x})^2\bar{q}\cdot P_1+ z (\frac{1-x}{1+x})(p_3-p_4)\cdot P_1-q\cdot P_1 }{\left(z^2(\frac{1-x}{1+x})\bar{q}\cdot P_1+ z (p_3-xp_4)\cdot P_1-x(\frac{1+x}{1-x})q\cdot P_1\right)(P_1 \leftrightarrow P_2)}\,.
\end{align}
This formula depends on the auxiliary variables $p_3$, $p_4$, $q$, and $\bar{q}$. It is easy to rewrite everything in terms of Mandelstam invariants and masses by first using \eqref{little} to find
\begin{align}
\label{p3-xp4}
(p_3 - x p_4)\cdot P_1 =& \frac{1}{2}\left(\frac{1+x}{1-x}\right)(\mb^2-\ma^2+s) \,, \\
(p_3 - x p_4)\cdot P_2 =& \frac{1}{2}\left(\frac{1+x}{1-x}\right)(\mb^2-\ma^2+u) \,
\end{align}
and then performing the scaling $z\rightarrow 2\frac{(1+x)}{M^2}\sqrt{-x}(q\cdot P_1) z$ to write the integral \eqref{mresult} in the compact form
\be
\label{compactint}
\frac{\mb}{\sqrt{-t}}\frac{M^4}{\left(4-\frac{t}{\mb^2}\right)^{\frac{5}{2}}}\frac{1}{2\pi i}\oint_{\Gamma} \frac{dz}{z^3}\frac{\left(\frac{1}{\sqrt{-x}}+\frac{(u-s)}{M^2}z+\sqrt{-x}z^2\right)^4}{\left(1-\frac{(\ma^2-\mb^2-s)}{M^2}
\frac{\sqrt{-t}}{\mb} z-z^2\right)
\left(1+\frac{(\ma^2-\mb^2-u)}{M^2}\frac{\sqrt{-t}}{\mb}z-z^2\right)}.
\ee
We have temporarily omitted the couplings. Recall that $M$ is defined in \eqref{M} as a solution of $M^4 = (\ma^2-\mb^2)^2-s u$ while $x$ is given by $(1+x)^2=t x/\mb$. Note that even though $M^2$ appears explicitly in the leading singularity contour integral, rescaling $z$ shows that it is only a function of $M^4$. The reason we keep it in the form given in \eqref{compactint} is to keep $z$ is dimensionless.

The branch of the square root in the solutions for $x$ can be changed by replacing $x\rightarrow \frac{1}{x}$ as it can be seen by writing the quadratic equation as $\frac{1}{\sqrt{-x}}-\sqrt{-x}=\sqrt{-t}/\mb$.

At this point we have to choose the contour $\Gamma$. Choosing the contour that computes residues at $z=0$ or $z=\infty$ gives rise to a triangle topology, while circling one of the two solutions of any of the two quadratic factors leads to box topologies. Let us choose the contour $\Gamma = S^1_{\infty}$ which computes the residue at $z=\infty$. This gives rise to the final form for $I_{-+}$.

One might have thought that since the residue is only a function of the Mandelstam variables then conjugating the internal helicity of the gravitons would have no effect on the answer and $I_{-+}$ would be equal to $I_{+-}$. This naive expectation is not true as the final answer is not a single valued function of $t$.

It can be shown that $I_{+-}$ can be obtained by performing the change $z\to - 1/z$ at the level of the integrand in $I_{-+}$. This in turn can be reabsorbed into a change of integration contour while keeping the integrand unchanged, effectively mapping $S^1_{\infty}\to - S^1_{0}$ (the minus sign coming from the inversion). This implies, as can be checked directly from \eqref{compactint}, that $I_{+-}$ corresponds to minus $I_{+-}$ evaluated on the other branch of $\sqrt{-t}$.  Alternatively, adding up both contributions one finds that $I_{-+}+I_{+-}$ can be written as \eqref{compactint} on the contour $\Gamma = S^1_{\infty}-S^1_{0}$.

Finally we move on to the remaining two helicity configurations. After inserting the four-point amplitude for same helicities \eqref{m4sameh} into the expression \eqref{gravitytriplecut}, we are left with a contour integral of the form
\be
\left(\dfrac{1-x}{1+x}\right) \int_{\Gamma_{\rm LS}}  \dfrac{zdz}{\left(z^2(\frac{1-x}{1+x})\bar{q}\cdot P_1+ z (p_3-xp_4)\cdot P_1-x(\frac{1+x}{1-x})q\cdot P_1\right)(P_1 \leftrightarrow P_2)}\,,
\ee
which has zero residue at both $z=0$ and $z=\infty$. Hence $I_{++}=I_{--}=0$ for the triangle leading singularity. This observation is consistent with results in the literature which use single unitarity cuts and also consider different helicity configurations in their computations \cite{Neill:2013wsa}.

Restoring the factors the gravitational coupling we define the leading singularity of the triangle topology, ${\rm LS}_{\rm triangle}/16\pi^2 G^2$, as \eqref{compactint} integrated on $\Gamma = S^1_{\infty}-S^1_{0}$ which is the result presented in \eqref{compactLS} in the introduction.

We now proceed to compute the box leading singularity. In contrast to the triangle, this will turn out to be analytic in $t$ and hence invariant under $x\rightarrow \frac{1}{x}$ . We can easily compute it by selecting one of the poles from the denominators, corresponding to the massive propagators of the scalar particle $\phi_A$. Let us now perform the cut in the propagator $(P_3+k_3(z))^2-\ma^2=2P_3\cdot k_3(z)$. Solving the quadratic equation for $z$, selecting one of the roots and computing the residue around it gives
\begin{align}
\label{exactbox}
{\rm LS_{\rm box}^{-+}} =& G^2 \pi^2 \frac{(s-\ma^2-\mb^2 \pm \sqrt{M^4-st})^4}{ t\sqrt{M^4-st}}\,\\
=&  G^2 \pi^2 \, \frac{\left(s-\ma^2-\mb^2 \pm \sqrt{(s-\ma^2-\mb^2)^2-4\ma^2\mb^2}\right)^4}{t\sqrt{(s-\ma^2-\mb^2)^2-4\ma^2\mb^2}}\,,
\end{align}
where the sign $\pm$ of the square root depends on the chosen root for $P_3\cdot k_3(z)=0$. The change in the sign can be shown to account for the parity flip, leading to the contribution ${\rm LS_{\rm box}^{+-}}$. This result turns out to be strikingly simple. Note that the denominator corresponds to the leading singularity of the scalar box \eqref{scalarboxls} while the numerator involves higher powers of the momenta, but tends to a constant in the non-relativistic limit $s\rightarrow (\ma+\mb)^2$. The fact that the gravitational leading singularity yields a pole $\frac{1}{t}$ enables us to easily extend the computation to $r$-loop ladder, as in the scalar case. This is discussed in section 4. Finally the case with equal helicities ${\rm LS_{\rm box}^{++}}$ can be treated analogously and yields
\begin{align}
\label{exactbox++}
{\rm LS_{\rm box}^{++}} 
=&  G^2 \pi^2 \, \frac{\left(\ma \mb \right)^4}{t\sqrt{(s-\ma^2-\mb^2)^2-4\ma^2\mb^2}}\,.
\end{align}

\section{Classical Gravitational Scattering of Two Massive Scalars}
\label{sec:dispersion}

In this section we aim to apply leading singularities to the computation of one-loop classical contributions to the scattering of two massive particles $A$ and $B$ with masses $\ma$ and $\mb$. This is a computation that has been performed in the literature using a variety of methods \cite{Blanchet:2002av,Futamase2007,Donoghue:2017pgk,Goldberger:2004jt,Holstein:08}. The techniques closest to our approach use on-shell methods such as BCFW recursion relations to efficiently compute tree-level amplitudes that are then used in unitarity cuts of loop amplitudes  \cite{Bjerrum-Bohr:2013,Neill:2013wsa,Holstein:2016fxh}. Using unitarity cuts for constraining, and sometimes completely determining, the integrand of an amplitude is known as the unitarity-based method developed mainly in the 90's for gauge theory computations \cite{Bern:1994cg,Bern:1994zx}. Once the integrand is known, reduction techniques are applied to write tensor integrals as sums over scalar integrals. The latter can be computed explicitly. In the non-relativistic limit two contributions are identified \cite{Holstein:08}, the first, usually denoted by $S = \pi/\sqrt{-t}$, leads to classical pieces and the second, $T = \log(-t)$, is quantum mechanical. As mentioned in the introduction, both contributions are generically present and are separated at the end of the computation.

The key idea in this section is to use a procedure we call multiple t-channel projections. In few words, we consider an amplitude as an analytic function of $t$ (possibly defined on a multi-sheeted Riemann surface with punctures). The projection corresponds to replacing the original function by one that agrees with the original on singularities at finite values of $t$ but which vanishes at infinity. In other words, we mod out by singularities at large $t$. This is analogous to what in dispersion relation theory are called subtraction terms. We find that at least up to one-loop, repeating this projection multiple times, in a way explained below, projects out quantum contributions and leaves behind the classical information.

In section 2 we explained how leading singularities are generalizations of standard unitarity cuts. The latter compute discontinuities of the amplitude across branch cuts. At one-loop, such discontinuities can themselves be functions with branch cuts. Leading singularities (LS) at one-loop compute such second discontinuities. This was explained in the case of a box-topology LS. As it turns out, the LS with a triangle topology computed in the previous section is the double discontinuity in the t-channel (The fact that the triangle topology LS has such an interpretation is discussed in detail in the next section). This means only the triangle topology survives the double t-channel projection and integrating back the double dispersion relation in the t-channel gives rise the classical contribution to the amplitude.

In section 5 we explain what higher loop generalizations of this construction and what the role of quantum corrections could be.

Before turning to the one-loop computation let us start at tree-level, which already motivates the idea of what we call the projection in the t-channel.

\subsection{Tree-Level Computation}

We use a similar parametrization of the momenta as in the previous section but now tailored to a BCFW computation that uses the t-channel.

Let $P_1=p_1+\x p_3$ and $P_3 = p_3+\y p_1$. Here $(p_1)_{\alpha,\dot\alpha}=\lambda_{1,\alpha}\tilde\lambda_{1,\dot\alpha}$ and $(p_3)_{\alpha,\dot\alpha}=\lambda_{3,\alpha}\tilde\lambda_{3,\dot\alpha}$ are null momenta while $\x=\ma^2/ 2p_1\cdot p_3$ and $\y = \mb^2/2p_1\cdot p_3$ ensure that $P_1^2=\ma^2$ and $P_3^2=\mb^2$. We proceed by using a BCFW deformation \cite{Britto:2004ap,Britto:2005fq} of the momenta via
\be
P_1(z) = P_1 + z\lambda_1\tilde\lambda_3, \quad P_3(z) = P_3 - z \lambda_1\tilde\lambda_3.
\ee
Clearly $P_1(z)^2=\ma^2$ and $P_3^2(z)=\mb^2$ for any value of $z$. The amplitude under consideration is $M^{\rm tree}_4(\{\Phi_A,P_1\},\{\Phi_A,P_2\},\{\Phi_B,P_3\},\{\Phi_B,P_4\})$ which under the deformation becomes a function of $z$ denoted $M_4(z)$. Here the subscript indicates the number of particles.

The BCFW construction starts with the identity
\be
M_4(0) = \frac{1}{2\pi i}\oint_{|z|=\epsilon} \frac{dz}{z}M_4(z).
\ee
Deforming the contour, or equivalently, using the residue theorem one finds an expression written in terms of the poles and residues of $M_4(z)/z$. There are two poles at finite locations in $z$. One is determined by requiring $t(z)=(P_1(z)+P_2)^2 = 0$ while the other by $u(z) =(P_1(z)+P_4)^2=0$. The pole at $u(z)=0$ has a non-zero residue only if there are interaction terms among particles $A$ and $B$. There is also a pole at $z=\infty$.

In preparation for the one-loop computation we write $t(z) = t + 2 z q\cdot P_2$. Recall that $q$ is the null vector $\lambda_1\tilde\lambda_3$. This means that
\be
M(0) = -\frac{1}{2\pi i}\oint_{|t(z)|=\epsilon} \frac{dz}{z} M(z) + \ldots = - \frac{1}{2\pi i}\oint_{|t'|=\epsilon} \frac{dt'}{t'-t}M(t')+\ldots
\ee
where the change of variables from $z$ to $t'=t(z)$ was performed. The ellipses in the first formula stand for other poles, either at finite locations or at infinity.

The explicit computation of the residue of $M(z)/z$ at $t(z)=0$ is very simple. Unitarity determines the residue to be
\be
M_3(P_1(z^*),P_2,G^-)\frac{1}{t}M_3(P_3(z^*),P_4,G^+)+M_3(P_1(z^*),P_2,G^+)\frac{1}{t}M_3(P_3(z^*),P_4,G^-)
\ee
where $z^*$ is satisfies $t(z^*)=0$.

Using the explicit form of the three-particle amplitudes the product can be written as (let us suppress the couplings temporarily)
\be
M_3(P_1(z^*),P_2,G^-)M_3(P_3(z^*),P_4,G^+) = \left(\frac{\langle k|P_1(z)|\tilde\eta]}{[k~\tilde\eta]}\right)^2\left( \frac{\langle \eta|P_3(z)|k]}{\langle \eta~k\rangle}\right)^2.
\ee
Here $\eta$ and $\tilde\eta$ are reference spinors that can be freely chosen while $k$ represents the momentum  vector of the graviton exchanged. A natural choice for the reference spinors is $\eta=\lambda_1$ and $\tilde\eta=\tilde\lambda_3$. The factors in the numerator simplify to $\langle k|P_1(z)|\tilde\lambda_3]= \langle k~1\rangle[1~3]$ and $\langle\lambda_1|P_3(z)|k]=\langle 1~3\rangle [3~k]$. Very nicely all dependence on $k$ cancels out to give
\be
M_3(P_1(z^*),P_2,G^-)M_3(P_3(z^*),P_4,G^+) = (2p_1\cdot p_3)^2 = \frac{\ma^2\mb^2}{\x\y}
\ee
where we have used $\langle 1~3\rangle [1~3]=2p_1\cdot p_3 $ and the definitions of $\x$ and $\y$.
A completely analogous computation shows that
\be
M_3(P_1(z^*),P_2,G^+)M_3(P_3(z^*),P_4,G^-) = (2p_1\cdot p_3)^2 (\x \y)^2 = \ma^2\mb^2 \x \y.
\ee
Combining both contributions one has that t-channel part of the amplitude is
\be\label{poles}
- \frac{1}{2\pi i}\oint_{|t'|=\epsilon} \frac{dt'}{t'-t}M(t') = \ma^2\mb^2\left(\x\y+\frac{1}{\x\y}\right)\frac{1}{t}.
\ee
Using $(P_1+P_3)^2 = (1+\x)(1+\y)2p_1\cdot p_3$ and the definitions of $\x$ and $\y$ it is easy to show that
\be
\label{M4tree}
M^{\rm tree}_4(\{\Phi_A,P_1\},\{\Phi_A,P_2\},\{\Phi_B,P_3\},\{\Phi_B,P_4\}) = \frac{(s-\ma^2-\mb^2)^2-2\ma^2\mb^2}{t} + \ldots
\ee
As the ellipses indicate, there are other pieces that are missing to obtain the full amplitude. However, as it is well-known, only the piece computed from the t-channel dispersion relation is needed in order to account for the long range interactions. The missing pieces are polynomials in $t$ and once the non-relativistic limit is taken these polynomials give rise to contact interactions.

We can now recover the standard form of the Newtonian potential in Fourier space by writing \eqref{M4tree} in the COM frame \cite{Holstein:08}, which is well suited to perform the non-relativistic expansion. In this frame
\begin{eqnarray}
t&=&-\vec{q}^{\,2} \,, \nonumber \\
s-\ma^2 - \mb^2&=& \left( 2\ma \mb +\frac{(\ma+\mb)^2}{\ma \mb}\vec{p}^{\,2}+ {\cal O}\left(|\vec{p}|^4\right)\right) +{\cal O}\left(|\vec{q}|^2\right) \,
\end{eqnarray}
where $\vec{q}$ corresponds to the momentum transfer vector\footnote{Here we slightly abuse the notation and use $\vec{q}$ for the three-momentum transfer while in other parts of the text we use $q$ for a reference massless four-vector. The meaning should be clear from the context and the use of the vector arrow in the case of the momentum transfer.}, and $\vec{p}$ accounts for the (average) momentum of the system. In this coordinates, the energies associated to each particle read
\be
E_{\textsc{a}}= \ma+ \frac{\vec{p}^{\,2}}{2m_{\textsc{a}}}+\ldots \,,\quad E_{\textsc{b}}= \mb+ \frac{\vec{p}^{\,2}}{2m_{\textsc{b}}}+\ldots \,
\ee
Restoring the couplings $\kappa = \sqrt{32 \pi G}$ in  \eqref{M4tree}, the classical potential is given by \cite{Feinberg:1988yw}
\be
\frac{M^{{\rm tree}}_{\rm classical}}{(4E_\textsc{A}E_\textsc{B})} = 4\pi G \frac{\ma \mb}{\vec{q}^{\,2}} \left(
1 + \frac{(3\ma^2 +8\ma \mb + 3\mb^2)}{2 \ma^2 \mb^2}\vec{p}^{\,2}+{\cal O}\left(|\vec{p}|^4\right)\right).
\ee

\subsection{One-Loop Computation}

The tree-level computation hints to the fact that dispersion relations in the t-channel are the only relevant ones to the classical scattering. In fact, one could have started from the trivial identity
\be\label{deform}
M_4^{\rm tree}(t) = \frac{1}{2\pi i}\oint_{|t'-t|=\epsilon}\frac{dt'}{t'-t}M_4^{\rm tree}(t')
\ee
and deformed the contour to pick up the pole at $t'=0$ and at $t'=\infty$. The former gives the result shown in \eqref{poles} while the latter gives the extra pieces not relevant for the long range interactions.

At one-loop level one expects to find branch cuts. Using the same dispersion relation formula \eqref{deform}, a contour deformation localizes the integral along a contour hugging the branch cut in the t-channel. The integral can be written as
\be\label{loop}
M^{\one}_4(t) = - \frac{1}{2\pi i}\int_{\Gamma_t}\frac{dt'}{t'-t}M^{\one}_4(t') + \ldots =-\frac{1}{2\pi i}\int_{t_i}^{t_f}\frac{dt'}{t'-t}\Delta_{t}M^{\one}_4(t') + \ldots.
\ee
where $\Gamma_t$ is the contour hugging the cut which starts at $t'=t_i$ and ends at $t'=t_f$. In the second equality the integral performed over the branch cut and the integrand is the difference of the values of $M^{\one}_4(t)$ on both sides of the cut at $t$. This is known as the discontinuity of $M^{\one}_4$ across the cut or $\Delta_{t}M^{\one}_4(t')$. Here again, the ellipses indicate other pieces which correspond to contributions from $t'=\infty$.

The presence of branch cuts seems to make the loop case different from the tree level one. However, this is not where the difference lies as the pole $1/t$ found at tree level can also be thought of as a branch cut by deforming it to $1/\sqrt{-t(4\mu^2-t)}$ with $\mu\ll 1$ an auxiliary mass scale. From this point of view the tree level formula is also given by \eqref{loop} with $t_i=0$ and $t_f=4\mu^2$ with the limit $\mu\to 0$ understood.

What makes one-loop level different from tree level is that $\Delta_{t}M_4^{\one}(t)$ has additional branch cuts. In fact it has another branch cut in the t-channel. Applying a dispersion relation argument again to $\Delta_{t}M_4^{\one}(t)$ one finds
\be\label{cutin}
M_4^{\one}(t) = \frac{1}{2\pi i}\int_{t_i}^{t_f}\frac{dt'}{t'-t}\frac{1}{2\pi i}\int_{t'_i}^{t'_f}\frac{dt''}{t''-t'}\Delta_{t}\Delta_{t}M_4^{\one}(t'') + \ldots.
\ee
It turns out that the double dispersion relation in this formula now contains all the classical scattering information. In other words, quantum corrections and contact interactions are projected away by the multiple discontinuities in the t-channel.

As discussed in the introduction of this section and in detail in the next section, the double discontinuity in the t-channel is nothing but the leading singularity associated with the triangle topology.

It is important to mention that we are treating the amplitude as an analytic function of $t,\ma$ and $\mb$ and not restricting it to a particular physical region. This means that even though we borrow the ``t-channel'' terminology, the computation is not restricted to that region. Therefore in addition to the standard branch cut used in dispersion relations in the t-channel which runs e.g. from $t_i=4\ma^2$ to $t_f=\infty$ \cite{Scherer:2002tk} we also allow another running from $t_i=-\infty$ to $t_f=0$. Once the contour of integration has been deformed to enclose both branch cuts we can equivalently express the result assuming that the branch cut runs from $t_i=0$ to $t_f=4\ma^2$ since we are ignoring terms that come from infinity. Of course, there is also a second branch cut now running from $t_i=0$ to $t_f=4\mb^2$.

The final formula for the one-loop contribution to the classical scattering is then
\be\label{cutoff}
M^{\one}_{\rm classical} = \frac{1}{(2\pi i)^2}\!\int_{0}^{4\ma^2}\!\!\!\frac{dt'}{t'-t}\int_{0}^{4\ma^2}\!\!\!\frac{dt''}{t''-t'}{\rm LS}_A(t'') + \frac{1}{(2\pi i)^2}\!\int_{0}^{4\mb^2}\!\!\!\frac{dt'}{t'-t}\int_{0}^{4\mb^2}\!\!\!\frac{dt''}{t''-t'}{\rm LS}_B(t'').
\ee

Let us rewrite the leading singularity ${\rm LS}_{B}(t)/16\pi^2 G^2$ computed in section \ref{sec:gravscat} for the reader's convenience
\be
\label{LSB}
\frac{\mb}{\sqrt{-t}}\frac{M^4}{\left(4-\frac{t}{\mb^2}\right)^{\frac{5}{2}}}\oint_{\Gamma} \frac{dz}{z^3}\frac{\left(\frac{1}{\sqrt{-x}}+\frac{(u-s)}{M^2}z+\sqrt{-x}z^2\right)^4}{\left(1-\frac{(\ma^2-\mb^2-s)}{M^2}
\frac{\sqrt{-t}}{\mb}z-z^2\right)
\left(1+\frac{(\ma^2-\mb^2-u)}{M^2}\frac{\sqrt{-t}}{\mb}z-z^2\right)}.
\ee
Recall that the contour $\Gamma = S^1_{\infty}-S^1_{0}$.
computes the residue at $z=\infty$ minus that at $z=0$ while $M$ and $x$ are defined via the equations
\be
M^4 = (\ma^2-\mb^2)^2-s u \quad {\rm and} \quad (1+x)^2=\frac{t}{\mb^2}x.
\ee
At this point one might worry that ${\rm LS}_B(t)$ is a very complicated function of $t$ and hence the dispersion integrals would lead to complicated functions. Moreover, the leading singularities are to be too singular around $t=4\mb^2$ to be integrable. Let us postpone the issue of the double pole at $t=4\mb^2$ and proceed to compute the residue at infinity and at $z=0$. The contour integral at $z=\infty$ has the form $(R_1 x+R_2 x^2+R_3 x^3)/M^4$ while that at $z=0$ is $-(R_1/x+R_2/x^2+R_3/x^3)/M^4$, where $R_a=R_a(s,u,\ma,\mb)$ are polynomials in their variables. Subtracting and multiplying by the prefactor in order to compute the leading singularity one finds
\be
{\rm LS}_{B}(t)=16\pi^2 G^2\frac{\mb}{\sqrt{-t}}\frac{M^4}{\left(4-\frac{t}{\mb^2}\right)^{\frac{5}{2}}}\left(R_1\left(x+\frac{1}{x}\right)+R_2 \left(x^2+\frac{1}{x^2}\right)+R_3 \left(x^3+\frac{1}{x^3}\right)\right).
\ee
It is easy to show that any combination of the form $x^m+x^{-m}$ is a polynomial in $t/\mb^2$ of degree $m$. For example, $x+1/x= (t/\mb^2)-2$.

Now we can go back to the problem of integrating this expression. Here we have to replace $u=-s-t+2(\ma^2+\mb^2)$ as all dependence on $t$ must be explicit for the integration. The polynomial of degree three in $t$ in the numerator now becomes one of degree six after removing $u$. The next step is to write this polynomial as one in $t-4\mb^2$. Finally, we are left with terms of the form
\be
\frac{1}{\sqrt{-t(4\mb^2-t)}}\frac{1}{(4\mb^2-t)^{2-m}} \quad {\rm with} \quad m\in \{0,1,2,\ldots,6\}.
\ee
When $m=0$ or $m=1$, the dispersion integral does not converge. The way to resolve the problem caused by the presence of the pole at $t=4\mb^2$ is to deform it so that it separates from the branch point by a small amount $\epsilon$. More explicitly, we compute
\be
\int_0^{4\mb^2}\frac{dt'}{(t'-t)\sqrt{-t'(4\mb^2-t')}}\frac{1}{(4\mb^2-t'+\epsilon )^{2-m}}.
\ee
The result of the integration can be expanded around $\epsilon =0$ to discover that all {\it singular} terms are again meromorphic functions in the t-complex plane with poles at $t=4\mb^2$, if any at all. These functions are of the same kind that the projection along the t-channel we defined above mods out by. The justification for doing so is that their final contribution to the amplitude only lead to terms that do not contribute to classical effects (see e.g. Appendix B in \cite{Bjerrum-Bohr:2013} for more details on why these kind of terms can be discarded).

Restricting our attention to the finite contributions in the $\epsilon$ expansion one discovers that, up to terms mod out by the t-projection, a copy of the original function, i.e.,
\be
\int_0^{4\mb^2}\frac{dt'}{(t'-t)\sqrt{-t'(4\mb^2-t')}}\frac{1}{(4\mb^2-t'+\epsilon )^{2-m}} = \frac{i\pi}{\sqrt{-t(4\mb^2-t)}}\frac{1}{(4\mb^2-t)^{2-m}}+\ldots.
\ee
In other words, these functions are self-similar under the dispersion relation projection. This means that the leading singularity itself can be taken to be self-similar. Therefore, a simple form of the classical contribution of the one-loop amplitude can be obtained
\be\label{oneloop}
M^{\one}_{\rm classical}(t) = \dfrac{{\rm LS}_A(t)+{\rm LS}_B(t)}{4}.
\ee
Restoring the couplings, the non-relativistic limit of \eqref{oneloop} in the center of mass frame is
\be
M^{\one}_{\rm classical}(t)= G^2 \pi^2 \frac{(\ma+\mb)}{|\vec{q}|}\left( 6 \ma^2\mb^2 +\frac{15}{2}(\ma+\mb)^2\vec{p}^{\,2}+{\cal O}\left(|\vec{p}|^4\right) \right) + {\cal O}\left(|\vec{q}|^0\right).
\ee
Therefore one obtains
\be\label{potentialOneLoop}
\frac{M^{\one}_{\rm classical}}{(4E_\textsc{A}E_\textsc{B})} =G^2 \pi^2  \frac{(\ma+\mb)}{|\vec{q}|}\left(6\ma \mb +\frac{(9(\ma^2+\mb^2)+30 \ma \mb)}{2\ma \mb}\vec{p}^{\,2} +{\cal O}\left(|\vec{p}|^4\right)\right).
\ee
which agrees with the results in the literature \cite{Neill:2013wsa}.

\section{More on Leading Singularities: Meaning and Higher Loops}
\label{discont}

In this section we collect some results on leading singularities which are either used in previous sections and require more detailed explanations or provide useful starting points for generalizations to higher loops.

\subsection{Triangle Leading Singularities as a Second Discontinuity}

One of the key ingredients in the previous section is the fact that the triangle topology leading singularity is the double discontinuity across the t-channel. In section 2 we briefly argued why the box topology leading singularity is the discontinuity in the t-channel of the function obtained by computing the discontinuity in the s-channel of a one-loop amplitude. In this section we give a more detailed explanation of this connection by working the cutting procedure taking into account the step functions involved.

Consider again the leading singularity for the scalar triangle of fig. \ref{fig:triangle1}b. Cutting the propagators $1/(L+P_3)^2$ and $1/(L-P_4)^2$ clearly computes the discontinuity in the t-channel. Further cutting $1/(L^2-m^2)$ and the emergent pole is very similar to the massless box computation of the first example. However, we will see that the meaning is very different: As anticipated, the second operation computes a discontinuity again across the t-channel.

\begin{figure}
  \centering
    \includegraphics[scale=0.30]{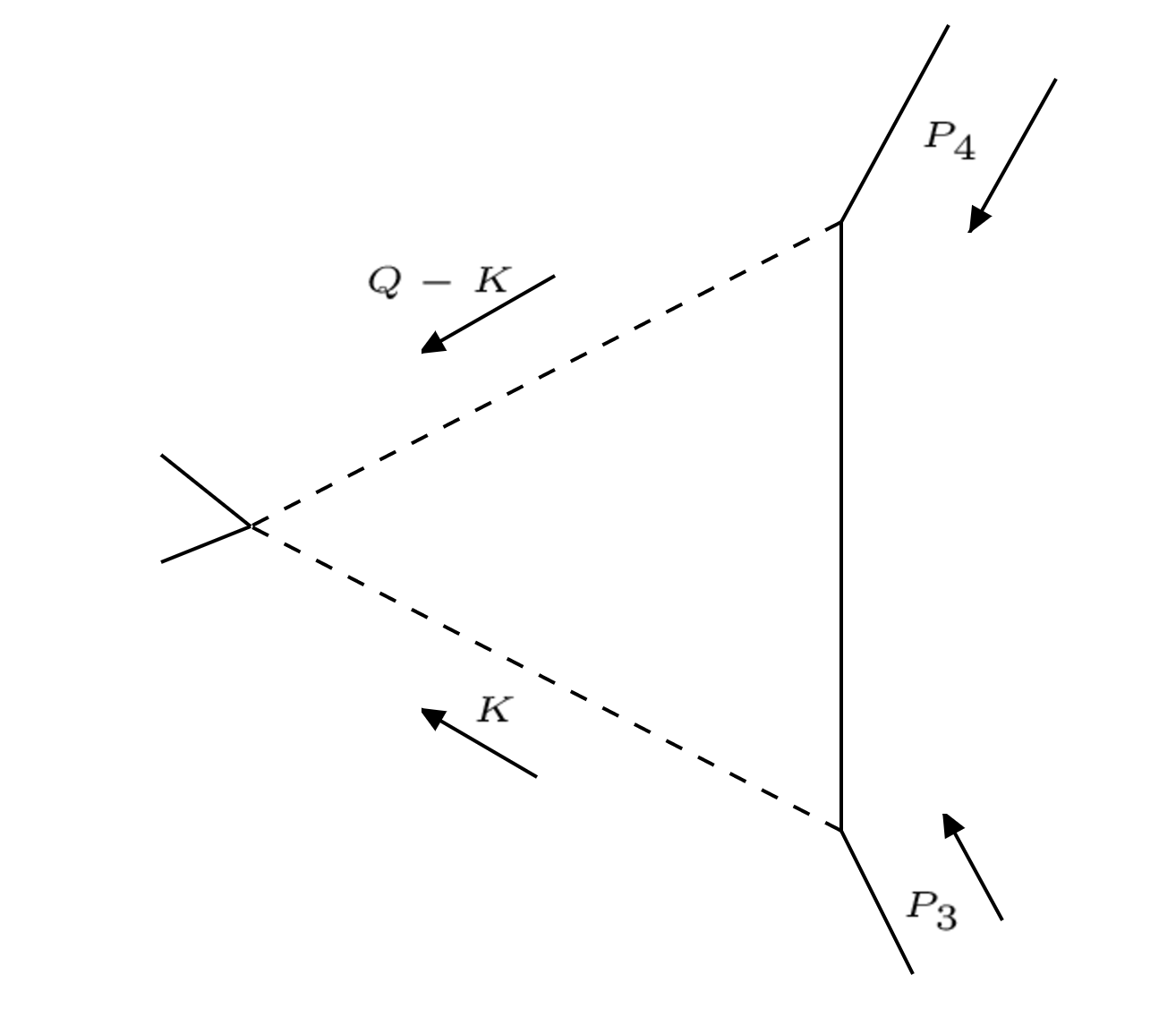}
     \caption{Alternative parametrization of the loop momenta in the scalar triangle diagram.}\label{scalartriangle}
\end{figure}

To illustrate this, let us compute explicitly the first unitarity cut. When regarded as a Feynman diagram, this corresponds to the imaginary part of the amplitude. This time let us parameterize the massless loop momenta as $K=v\lambda \tilde{\lambda}$. Define also $Q:=P_3+P_4$ to be the momentum transfer. The discontinuity can then be written as \cite{Britto:2004nc}
\be
\label{firstdisc}
\Delta_{\rm triangle}=-\frac{1}{4(2\pi)^2 }\int_{\Gamma_{\Delta}} \dfrac{ dv \langle \lambda ~d\lambda \rangle [\tilde\lambda ~d\tilde\lambda]}{[\tilde\lambda |P_3|\lambda\rangle }\delta(Q^2+v[\tilde\lambda |Q|\lambda\rangle) \,,
\ee
where the delta function arises from cutting the second massless propagator, with momentum $Q-K$, see fig \ref{scalartriangle}. Here the contour $\Gamma_{\Delta}$ is defined such that the loop momenta $K$ is real, i.e. $v \in \mathbb{R}$ and $\lambda^{\dagger}=\tilde{\lambda}$. After fixing $v=-\frac{Q^2}{[\tilde\lambda |Q|\lambda\rangle}$ we find that the integrand develops a new pole in $[\lambda |Q|\lambda\rangle$:
\begin{align}
\Delta_{\rm triangle}=&-\frac{1}{4(2\pi)^2 }\int_{\Gamma_{\Delta}} \dfrac{  \langle \lambda~ d\lambda \rangle [\tilde\lambda ~d\tilde\lambda]}{[\tilde\lambda |P_3|\lambda\rangle [\tilde\lambda |Q|\lambda\rangle}\,.
\end{align}
The meaning of new the pole is clear: It corresponds to $v=-\frac{Q^2}{[\tilde\lambda |Q|\lambda\rangle}\rightarrow \infty$, rendering the loop momenta $K$ divergent. This is exactly what we get for $z\rightarrow 0$ or $z\rightarrow \infty$ in the previous parametrization (see \eqref{k3k4}). Thus we again find the existence of a hidden pole in the triangle, arising from the measure of the cut in the visible propagators.

We can solve the integral for the spinor helicity variables by introducing Feynman parameters, and then performing the integral over the real contour $\Gamma_{\Delta}$ as in \cite{Cachazo:2004kj}
\begin{align}
\Delta_{\rm triangle}=&
-\frac{1}{4(2\pi)^2 } \int_{0}^1 du \int_{\Gamma_{\Delta}} \dfrac{  \langle \lambda ~d\lambda \rangle [\tilde\lambda ~d\tilde\lambda]}{[\tilde\lambda |(1-u) P_3+uQ|\lambda\rangle^2} \,\nonumber \\
=&\frac{1}{4(2\pi i ) }  \int_{0}^1 du  \dfrac{ 1}{((1-u) P_3+uQ)^2} \,\nonumber \\
=&\frac{1}{4(2\pi i ) } \int_{0}^1 du  \dfrac{ 1}{(1-u)^2\mb^2 + u t} \,\nonumber\\
=&\dfrac{1}{4\mb^2} \frac{1}{2\pi i } \int_{0}^1 du  \dfrac{ 1}{(u-u_{+})(u-u_{-})}\,,
\end{align}
where $u_{+}=-x$ and $u_{-}=-\frac{1}{x}$, and $x$ is given by \eqref{xandt}. At this stage we are interested in the analytic properties of $\Delta_{\rm triangle}$, hence we regard it as a function of the complexified $t$ variable, with potential branch cuts. We assume that the solutions $u_{+}$ and $u_{-}$, for a given $t \in \mathbb{C}$ do not lie in the interval $[0,1]$, such that the integral converges. With this considerations, the expression can be explicitly integrated to give
\be
\Delta_{\rm triangle} = \frac{1}{4 \pi \sqrt{(-t)(4\mb^2-t)}}{\rm arctanh}\left(\sqrt{\frac{4\mb^2-t}{-t}}\right).
\ee
We can now study the behavior of $\Delta_{\rm triangle}$ associated to analytically continuing $t\rightarrow e^{i\phi}t$, with $0\leq \phi < 2\pi$. For  $t \in (0,4\mb^2)$ we find a further discontinuity corresponding to the square root factor. This is easily seen by noting that
\be
u_{+}-u_{-}= \dfrac{1-x^2}{x}=\dfrac{\sqrt{(-t)(4\mb^2-t)}}{\mb^2}\,.
\ee
Hence, for  $t \in (0,4\mb^2)$, the complex rotation corresponds to exchanging the roots $u_-$ and $u_+$.  The discontinuity associated to the exchange of the roots of a second order polynomial can be easily computed, by contour deformation, as the residue in any of such roots. We thus have
\begin{align}
\label{2nddisc}
\Delta^2 _{\rm triangle}=& \dfrac{1}{4\mb^2} \dfrac{ 1}{(u_{+}-u_{-})} \Theta(t) \Theta(4\mb^2 - t )\, \nonumber \\=& \dfrac{ 1}{4 \sqrt{(-t)(4\mb^2-t)}} \Theta(t) \Theta(4\mb^2 - t ) \,.
\end{align}
Note that here we have explicitly written the step functions, which account for the location of the new branch cut. The process can be trivially iterated to compute further discontinuities, leading to the exact same contribution \eqref{2nddisc} (up to factors of $2$), which can be understood as the ``maximal ambiguity" in the $t$ channel. That is to say, the expression is self similar under higher dispersion relations in the sense explained in section \ref{sec:dispersion}. The location of the new branch cuts coincide with the original one coming from the first discontinuity, that is, the unitarity cut \cite{analytic}. The two branch points $t=0$, $t=4\mb^2$ have the physical meaning of being the threshold for production of massless and massive states, respectively \cite{Holstein:2004dn}. Furthermore, they correspond to the non-relativistic limit of different physical regions, i.e. physical processes in the $t$ and $s$ channel, respectively \cite{Donoghue:1996kw,Scherer:2002tk}.

The main observation is that \eqref{2nddisc} is precisely the leading singularity computed in subsection \ref{scalar}. In fact, computing the residue in the $u$-plane accounts for cutting the massive propagator $[\tilde\lambda|P_3|\lambda\rangle$ in the integral over $\lambda$, $\tilde \lambda$. The real contour $\tilde{\lambda}=\lambda^{\dagger}$ can then be deformed to circle the emergent pole $[\tilde\lambda|Q|\lambda\rangle=0$.

We can now provide further details of this process for the case of the complete one-loop gravitational scattering represented in fig. \ref{blob}. It is instructive to consider complexified external momenta, such that the momentum transfer reads $Q=P_3+P_4=(iq,0,0,0)$, with $t=Q^2=-q^2$. For this configuration, all the massive propagators $[\tilde\lambda|P_i|\lambda\rangle$, $i=1,2,3,4$ have support on the real contour given by $\Gamma_{\Delta}$. Specifically, the first discontinuity of the 1-loop amplitude has the form
\begin{equation}
\Delta_{\rm full} = \frac{1}{Q^2}\int_{\Gamma_{\Delta}}  \langle \lambda ~d\lambda \rangle [\tilde\lambda ~d\tilde\lambda] \dfrac{F(\lambda,\tilde{\lambda},P_1,P_3,Q)}{[\tilde\lambda|Q|\lambda\rangle ^6\prod_{i=1}^4([\tilde\lambda|P_i|\lambda\rangle+i\epsilon[\tilde\lambda|Q|\lambda\rangle)}
\end{equation}
where $F$ is a polynomial in its arguments and $\Gamma_{\Delta}$ is defined by $\tilde{\lambda}=\lambda^{\dagger}$. We have also absorbed a factor of $Q^2$ into the definition of $\epsilon$. $\Gamma_{\Delta}$ can be easily parameterized by putting $\lambda=(1,x+iy)$, with $\langle \lambda~ d\lambda \rangle [\tilde\lambda ~d\tilde\lambda]=dx dy$. The second discontinuity can then be understood as the ambiguity in the $i\epsilon$ prescription, which yields the corresponding ambiguity in $\Delta_{\rm full}$ as a function of $t$. In order to compute it, it is natural to cut the massive propagators one by one, which already induces the triangle cut of figure \ref{fig:gravtrian}, and sum over all such residues. The remaining 1-dimensional contour in the $x,y$ variables can be deformed to encircle the pole at $[\tilde\lambda|Q|\lambda \rangle$, yielding the triangle leading singularity.

We have also shown that the triangle leading singularity encodes the precise non-analytical structure needed to recover the 1-loop effective potential from gravitational scattering. This turns it into a natural candidate for evaluating classical corrections to low energy phenomena in a wide range of effective field theories.

The simplicity of the leading singularity computation, as contrasted with previous approaches, strongly motivates the study of higher loop corrections to long range interactions. This is further supported by the fact that these quantities do not suffer from divergences which are common in loop integrals, and hence become good candidates for building blocks of a low energy energy effective theory. Also, as seen in section \ref{scalar}, the 1-loop box and triangle diagrams define a contour that certainly projects out all the other scalar integrals. Thus, the leading singularity contour can be used at higher loops to compute coefficients in the scalar integral expansion, in order to decide if a given scalar diagram contributes to the classical potential. We now proceed to point out some progress in these directions.

\subsection{Higher-Loop Examples}

\subsubsection{Iterating Triangle Leading Singularities}
\label{iteratingtrian}

\begin{figure}
  \centering
    \includegraphics[scale=0.30]{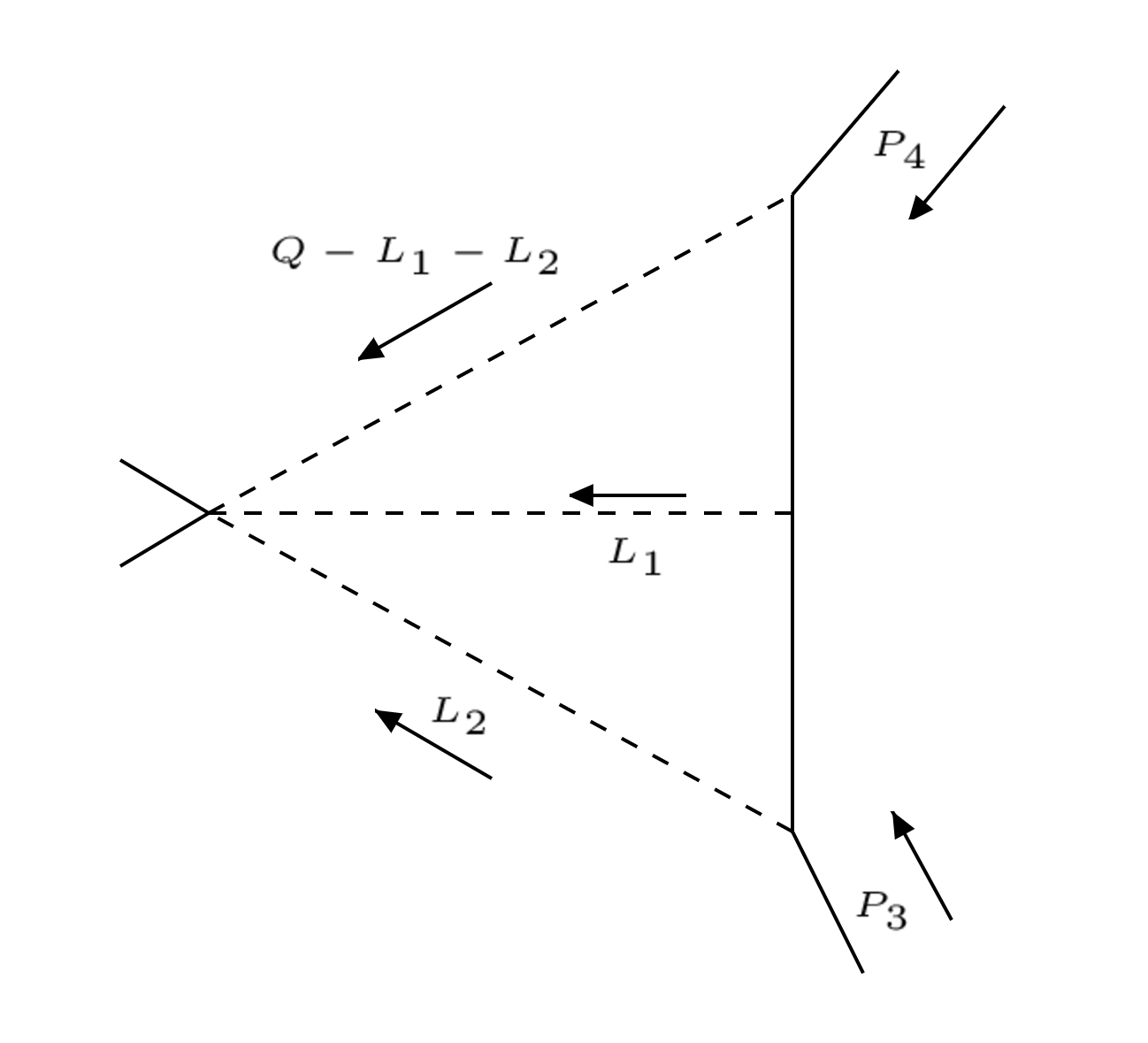}
   \caption{$2$-loop nested triangle. The solid lines represent massive particles exchanging massless states, represented by dashed lines.}\label{double}
\end{figure}

We can iterate the result for the scalar triangle to an arbitrary number of loops. Consider first the double triangle diagram of fig. \ref{double}, where the five visible propagators are cut. All the particles in the diagram are on-shell, thus we can first compute the leading singularity in the upper triangle. As shown in the previous subsection, performing the triple cut in the upper triangle and taking the residue at $z=\infty$ is equivalent to introduce Feynman parameters and compute the residue in the $u$ plane. The result is given by \eqref{2nddisc}, which we can now write as
\begin{equation}
\label{upperls}
\frac{1}{4(2\pi i ) } \int_{\Gamma_{\rm LS^1}} du  \dfrac{ 1}{(1-u)^2\mb^2 + u \hat{Q}^2}= \dfrac{ 1}{4 \sqrt{(-\hat{Q}^2)(4\mb^2-\hat{Q}^2)}} \,.
\end{equation}
Here $\hat{Q}=Q-L_2$, and we have omitted the functions  $\Theta(\hat{Q}^2) \Theta(4\mb^2-\hat{Q}^2 )$ associated to the second branch cut. We will now see, however, that these precisely define the corresponding contour ${\rm LS^2}$ for the Leading Singularity. Parameterizing $L_2=v \lambda \tilde{\lambda}$, the leading singularity of the full diagram reads
\be
\label{branchls}
{\rm LS}=\int_{\Gamma_{\rm LS^2}} \dfrac{ dv \langle \lambda ~d\lambda \rangle [\tilde\lambda~ d\tilde\lambda]}{[\tilde\lambda |P_3|\lambda\rangle } \dfrac{ 1}{\sqrt{(-\hat{Q}^2)(4\mb^2-\hat{Q}^2)}}\,
\ee
where $\hat{Q}^2(v)= Q^2 + v [\tilde\lambda | Q| \lambda \rangle$. We find the presence of a branch cut in the $v$-plane directly arising from the integration of the upper triangle. This is the analog of the hidden pole $[\tilde\lambda|Q|\lambda\rangle$ that appears in the 1-loop case, thus it is natural to define the new contour ${\rm LS^2}$ to enclose it. Furthermore, this accounts for inserting the function $\Theta(\hat{Q}^2) \Theta(4\mb^2-\hat{Q}^2 )$ into the integrand, which arise naturally as part of the second discontinuity \eqref{2nddisc}. The integration over the branch cut in $v$ is easily done after putting $v=-\frac{Q^2}{ [\tilde\lambda | Q| \lambda \rangle}y$ and yields eq. \eqref{finaldiscont} below. However, let us retrace our steps to see if there is another way leading to the final result. Let us write the first leading singularity using the LHS of equation \eqref{upperls} and commute the integration over the $u$ and $v$ variables in \eqref{branchls}. This leads to
\be
{\rm LS}= \int_{} \dfrac{ \langle \lambda ~d\lambda \rangle [\tilde\lambda~ d\tilde\lambda]}{[\tilde\lambda |P_3|\lambda\rangle[\tilde\lambda |Q| \lambda \rangle }  \int_{} du \int dy  \dfrac{ Q^2}{(1-u)^2\mb^2 + u (1-y) Q^2}\,.
\ee
Again we have set $v=-\frac{Q^2}{ [\tilde\lambda | Q| \lambda \rangle}y$. In the integral sign we have omitted the explicit contours for simplicity. We find the emergent pole $[\tilde\lambda |Q| \lambda \rangle$ arising from the Jacobian of the $v$-integration. This time, however, we note that there is no branch cut in the $y$ plane, but instead a simple pole, which residue readily gives
\be
\label{finaldiscont}
{\rm LS}=\int_{\Gamma_{\rm LS}} \dfrac{ \langle \lambda ~d\lambda \rangle [\tilde\lambda ~d\tilde\lambda]}{[\tilde\lambda |P_3|\lambda\rangle [\tilde\lambda | Q| \lambda \rangle} \int \dfrac{du}{u}
\ee
which again generates a simple pole in the $u$ plane, nicely turning the original quadratic denominator of \eqref{upperls} into a linear one. After restoring the corresponding factors, the residue in the $u$ plane is simply

\be
\label{finaldiscont}
{\rm LS}=-\dfrac{1}{(2\pi)^2}\int_{\Gamma_{\rm LS}} \dfrac{ \langle \lambda~ d\lambda \rangle [\tilde\lambda~ d\tilde\lambda]}{[\tilde\lambda |P_3|\lambda\rangle [\tilde\lambda | Q| \lambda \rangle}= \dfrac{ 1}{4 \sqrt{(-t)(4\mb^2-t)}}.
\ee

We stress that this is exactly the same result as if the integration over the branch cut in \eqref{branchls} had been performed. This fact reflects the intrinsic nature of the leading singularity and its defining contour. More precisely, it provides evidence that the operation here implemented neither depends on the order of integration nor the parametrization used.

Finally, in this way we can continue to iterate the result to any number of triangles arranged in the nested topology of fig \ref{double}. We conclude that the leading singularity reflects the non-analytical structure proper to the triangle topology, which has been extensively discussed in \cite{Holstein:2004dn} for the 1-loop case.

\subsubsection{A 2-loop Example for Gravity}

Here we demonstrate that leading singularities can be used at higher loops to evaluate the contribution from a given scalar integral. At two loops, the scalar diagrams associated to the classical potential have been surveyed in \cite{Gilmore:2008gq} in  the context of Non-Relativistic GR (NRGR). We can study the leading singularities associated to each of these diagrams, and use them to define the integration contours $\Gamma_{{\rm LS}}$. These contours can then be implemented in the  2-loop gravity amplitude, and the result can be used to obtain the respective coefficients in the scalar expansion.

\begin{figure}
\centering
 \subfloat[]{{\includegraphics[width=7cm]{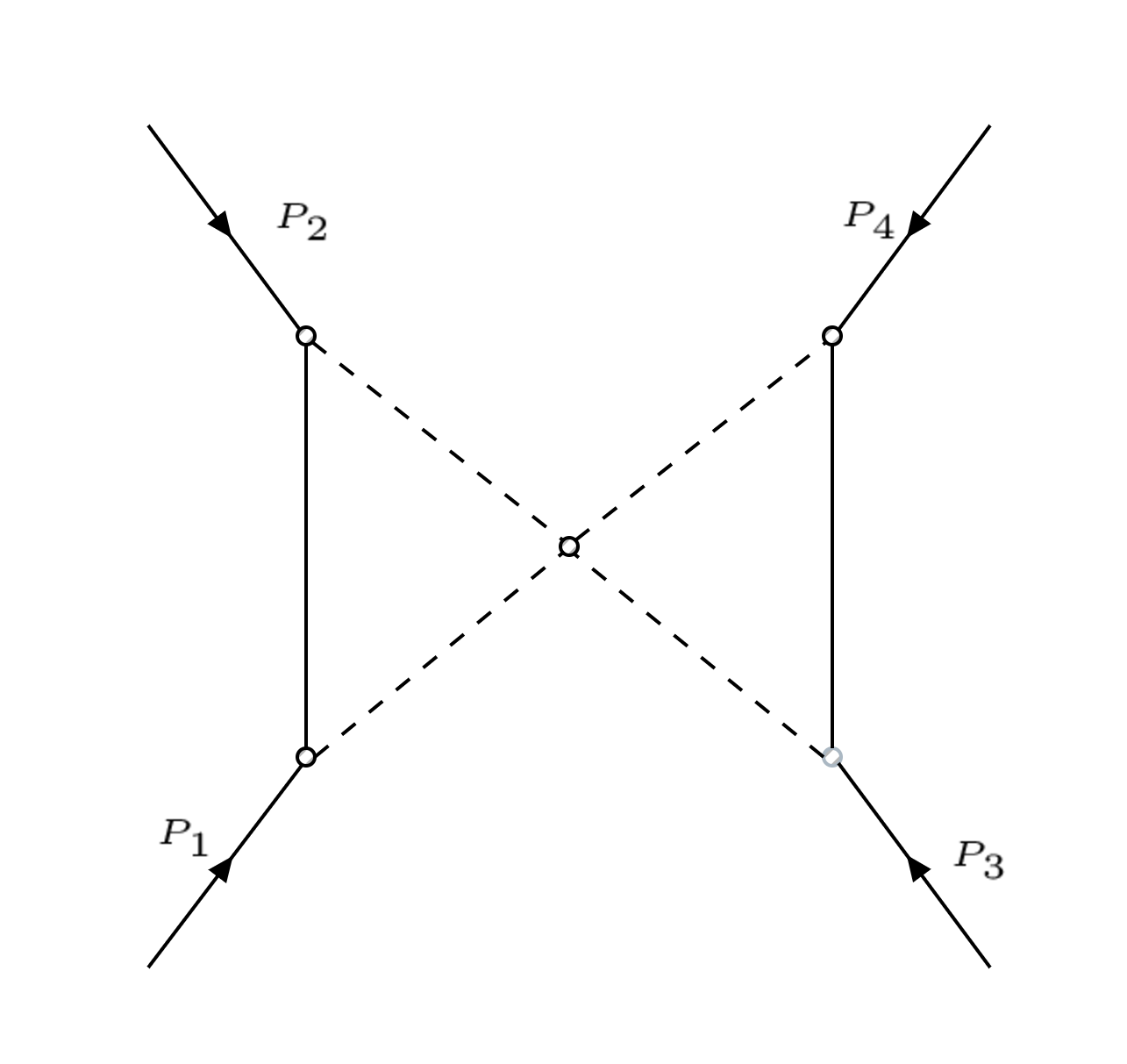} }}%
    \qquad
        \subfloat[]{{\includegraphics[width=7cm]{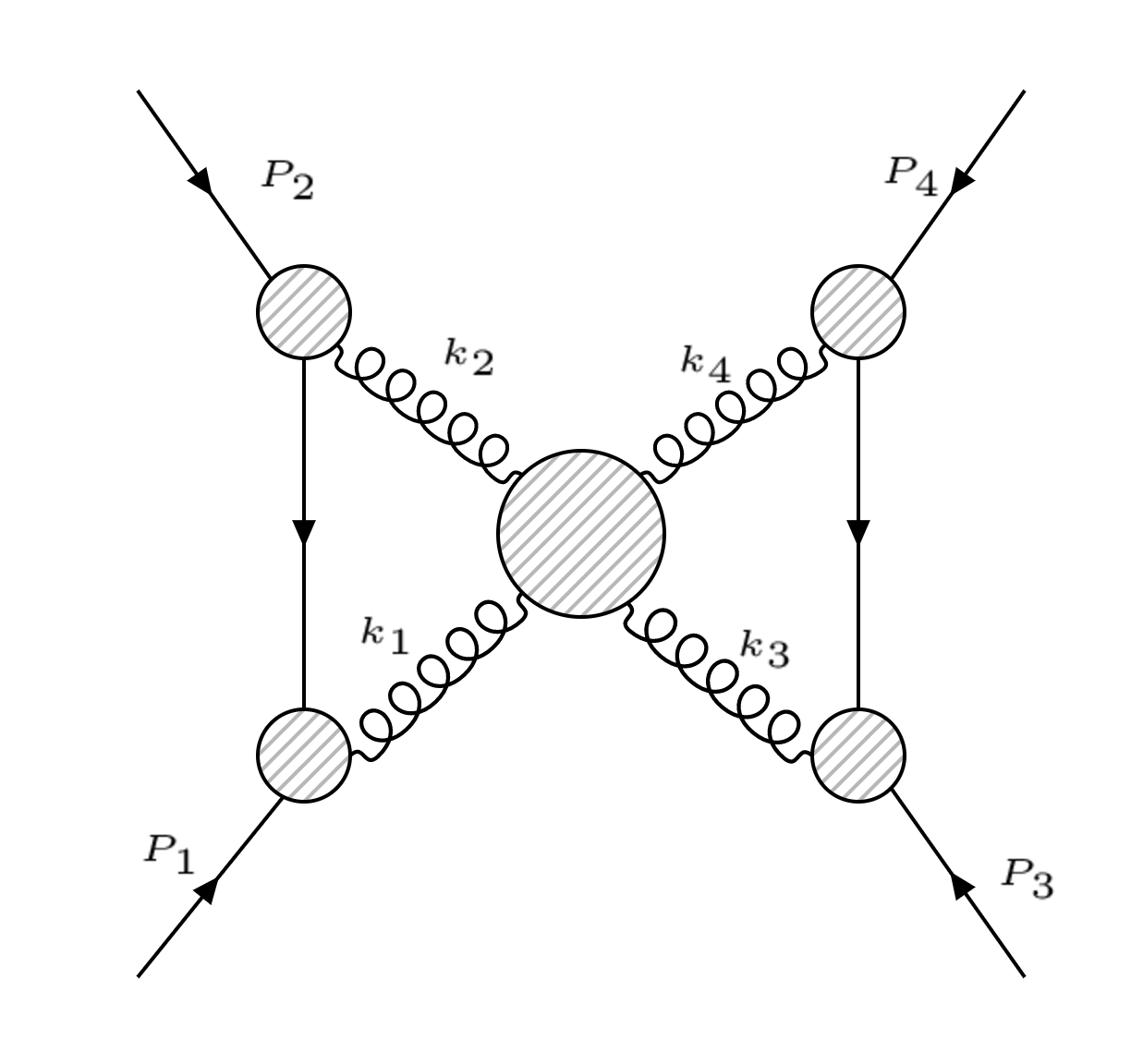} }}%
    \caption{The product triangle scalar diagram and its gravitational counterpart. The gravity diagram also contains double boxes that emerge when one of the propagators is cut in the 4-graviton amplitude.}%
    \label{fig:2looptrian}
\end{figure}

We consider only the simplest example, a complete analysis being left for future work. We evaluate the product triangle of Fig. \ref{fig:2looptrian}a and argue that its contribution to the gravitational amplitude is quantum mechanical in nature. This is consistent with the results found in \cite{Gilmore:2008gq} using NRGR.

In this case $\Gamma_{{\rm LS}}= \Gamma_{(S^1)_{\infty}\times (S^1)_{\infty}}$, and the leading singularity trivially gives (up to irrelevant factors)
\be
{\rm LS}_{pt} = \dfrac{ 1}{t \sqrt{(4\ma^2-t)} \sqrt{(4\mb^2-t)}}.
\ee
On the other hand, the full contribution from the triangle scalar integral of Fig. \ref{scalartriangle} can be found in Eq. B.5 of \cite{Bjerrum-Bohr:2016hpa}. This result can be used to evaluate the full contribution of the product triangle. Here we just provide the non-relativistic limit which will determine the long range behavior of the potential. Up to irrelevant factors, the leading terms for $t\rightarrow 0$ read
\be
\label{Ipt}
I_{pt}=-\frac{ \ma \mb}{t}+\frac{\ma+\mb}{\pi^2\sqrt{-t}} {\rm log}\left( \frac{-t}{\ma \mb}\right)+\frac{1}{\pi^4} {\rm log}^2\left( \frac{-t}{\ma \mb}\right)+\ldots
\ee

We note the presence of $t^{-1}$ as leading term in the expansion. This may seem puzzling at first sight, since after Fourier transformation this term leads back to the Newtonian potential \cite{Donoghue:1994dn}. However, such contribution yields in fact a contact term in the gravitational potential. To see this, recall that in the amplitude the contribution $I_{pt}$ is multiplied by the corresponding coefficient $c_{pt}$. Assuming the contour $\Gamma_{S^{\infty}\times S^{\infty}}$ projects out all the other scalar integrals, we find
\be
c_{pt}=\frac{{\rm LS}^{\rm grav}_{pt}}{{\rm LS}_{pt}}\,.
\ee

In order to compute ${\rm LS}^{\rm grav}_{pt}$ we use the expression \eqref{mresult}, and insert the 4-graviton amplitude $M_4(k_1^{-},k_2^{+},k_3^{-},k_4^{+})$ instead of the Compton amplitude $M_4(P_1,P_2,k_3^{-},k_4^{+})$. We then consider a copy of this integral associated to the particle $m_a$. The full expression now reads
\begin{eqnarray}
\label{mresult2}
{\rm LS}^{\rm grav}_{pt}&=&\dfrac{\kappa^4 \ma^2\mb^2(xy)^3}{(1-x^2)(1-y^2)}\left(\dfrac{1+x}{1-x}\right)^4 \left(\dfrac{1+y}{1-y}\right)^4 \nonumber \\
&&\times \left(\frac{1}{(2\pi i)^2 }\int_{\Gamma_{\rm (S^1)_{\infty}\times (S^1)_{\infty}}} \dfrac{dz}{z^5}\dfrac{d\omega}{\omega^5}M_4(k_1^{-},k_2^{+},k_3^{-},k_4^{+}) \right)\,,
\end{eqnarray}
where $y$ is defined as in \eqref{p3-xp4} for $P_1$ and $P_2$, and $k_1$,$k_2$ are defined as in \eqref{k3k4}. The 4-graviton amplitude is given by
\be
M_4(k_1^{-},k_2^{+},k_3^{-},k_4^{+}) = \frac{\kappa^2}{t}\times \dfrac{[k_4~k_2]^4\langle k_1~k_3\rangle ^4}{[k_3 ~k_1]\langle k_3~k_1\rangle[k_3~k_2]\langle k_3~k_2\rangle}\,.
\ee

Again, it is easy to check that the configurations $h_1=h_2$, $h_3=h_4$ lead to vanishing residue in $\Gamma_{\rm S^{\infty}\times S^{\infty}}$. After performing the change of variables $z=\left(\frac{1+x}{1-x}\right)u$, $\omega=\left(\frac{1+y}{1-y}\right)v$, the residue can be computed exactly. The coefficient then takes the form
\be
c_{pt}=t \times \kappa^6 (\ma \mb)^4 P_2\left( \frac{s-\ma^2 -\mb^2}{\ma \mb}\right)  + O(t^2)\,,
\ee
where $P_2$ is a second order polynomial. Hence, we conclude that the leading term in \eqref{Ipt} becomes a contact term when multiplied by $c_{pt}$. Having shown that the leading contribution is quantum mechanical, it is easily argued that the subleading terms are of the same nature. In fact, as argued in \cite{Holstein:2004dn}, the $\hbar$ factors can be restored in the result by keeping track of the combination $\frac{m}{\sqrt{-t}}\rightarrow \frac{1}{\hbar} \frac{m}{ |\vec{k}|}$. We conclude that the expansion in \eqref{Ipt} precisely corresponds to an expansion in $\hbar$, hence the full contribution of $I_{pt}$ is quantum in nature.

\subsubsection{Infinite Family of Leading Singularities}
\label{inf}

In this part we present an all-loop iterative result for ladder diagrams. The construction is similar in nature to that presented above in iterated triangle case. However, in this case it can be readily extended to its gravitational counterpart. Ladder diagrams have also been studied in the context of NRGR, see e.g. \cite{Goldberger:2007hy}.

\begin{figure}
  \centering
    \includegraphics[scale=0.40]{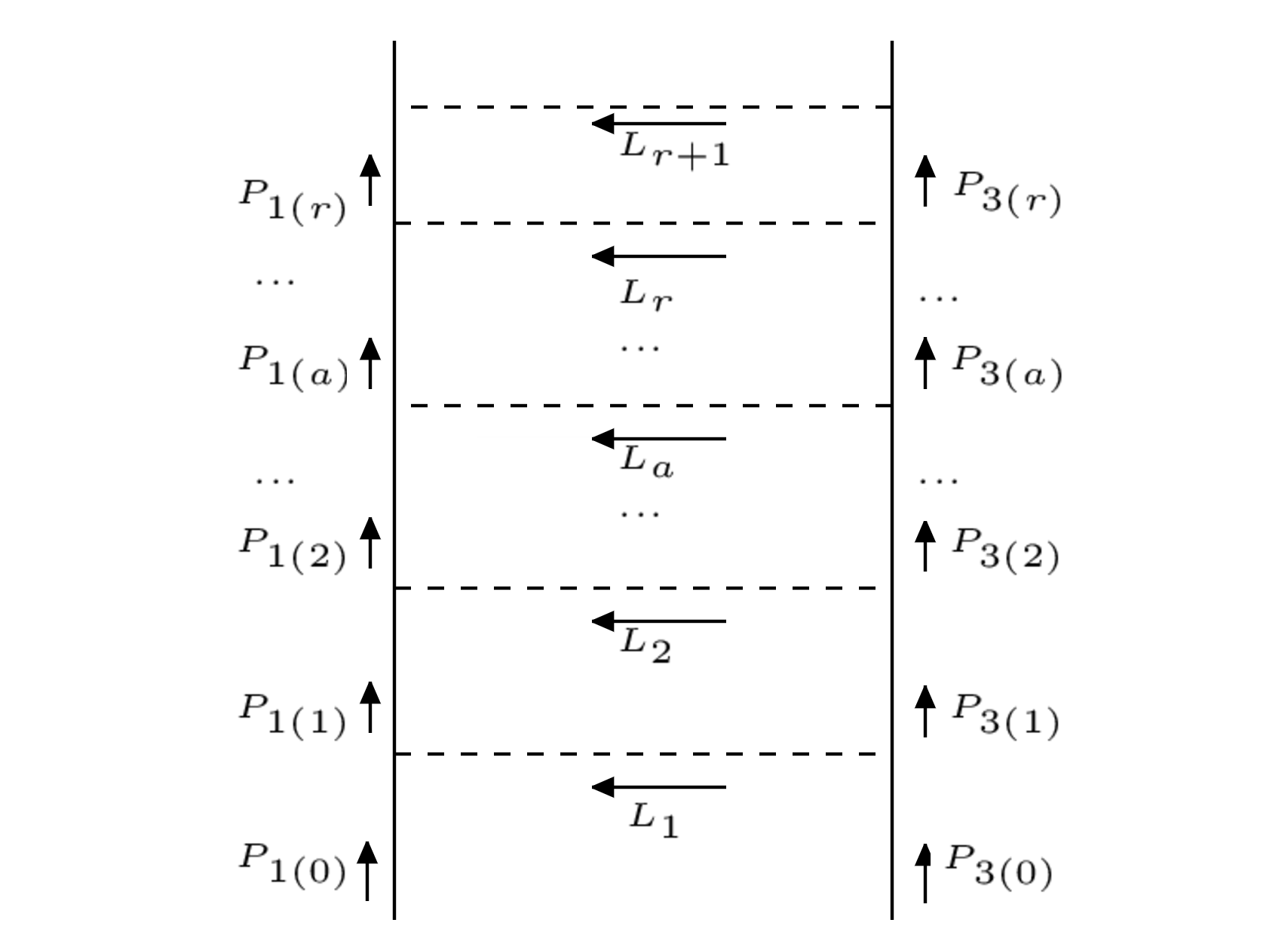}
\caption{$r$-loop ladder. The solid lines represent two different massive particles exchanging $r+1$ massless states, represented by dashed lines.}\label{rbox}
\end{figure}

Consider the case in which both massive scalar can only interact with the massless scalars via three particle couplings. We would like to consider a family of leading singularities with the topology of a ladder with $r+1$ rungs, i.e., with $r$ loops, as in fig. \ref{rbox}. In the formula we will suppress the three-particle amplitudes as they are all given by the coupling constant. The integral to be performed is then
\be
\int \prod_{a=1}^r d^4L_a \left(\prod_{a=1}^{r-1}\frac{1}{L_a^2(P_{1(a)}^2-\ma^2)(P_{3(a)}^2-\mb^2)}\right)
\frac{1}{L_m^2(P_{1(m)}^2-\ma^2)(P_{3(m)}^2-\mb^2)L_{m+1}^2}
\ee
where momenta are defined recursively as follows $P_{1(a)} = P_{1(a-1)} + L_a$, $P_{3(a)} = P_{3(a-1)} - L_a$ with $P_{1(0)} = P_1$, $P_{3(0)} =P_3$ and $L_{r+1} = P_3+P_4-(L_1+L_2 \cdots + L_{r-1})$.

Once again we seem to be in the situation of having to find $4r$ poles while given only $3r+1$ propagators. As in the previous case, emergent propagators account for the difference.

Starting with the integral over $L_r$ one has a box integral with external particles on-shell (as the contour encloses the propagators $1/(P_{1(r-1)}-\ma^2)$ and $1/(P_{3(r-1)}^2-\mb^2)$). This box contour integral is easy to compute and gives
\be
\label{itbox}
\frac{1}{\sqrt{(s-\ma^2-\mb^2)^2-4\ma^2\mb^2}}\times \frac{1}{(P_3+P_4-(L_1+L_2 \cdots + L_{r-2}))^2}.
\ee
Note the appearance of the $1/(P_3+P_4-(L_1+L_2 \cdots + L_{r-1}))^2$. This is nothing but an emergent propagator which happens to be exactly the one needed to complete the left over integrations into an $r-1$ loop ladder integral. The procedure can be iterated until all integrals are completed and the result is
\be
{\rm LS}^r_{{\rm ladder}}=\frac{1}{((s-\ma^2-\mb^2)^2-4\ma^2\mb^2)^{r/2}t}
\ee
where again $t=(P_3+P_4)^2$.

Using the exact result of section \ref{sec:gravscat}, we can repeat the construction for the gravitational case. Let us now sketch the differences in the procedure. In the gravitational case the vertices in fig. \ref{rbox} correspond to the on-shell 3pt amplitudes $M_3(\Phi_A,\Phi_A,G)$, the internal lines being gravitons. Consider, for instance, the helicities of the gravitons to be $h_i=-2$, except for $h_{r+1}=+2$. After cutting the top box, that is, the integral over $L_r$, we end up with the same expression \eqref{itbox}, now multiplied by the numerator of \eqref{exactbox}. We again use the emergent propagator to close the remaining ladder. However, this time we need to also restore the 3pt amplitudes given by
\be
M(P_{1(r-1)},P_2,G^{\pm}\!)M(P_{3(r-1)},P_4,G^{\mp}\!)\! = \!\left(s-\ma^2-\mb^2 \pm \sqrt{(s-\ma^2-\mb^2)^2-4\ma^2\mb^2}\right)^2\!\!,
\ee
which means we need to extract this factor from the result of the first integration and attach it to the next one. After iterating this procedure $r$ times we will get
\be
{\rm LS}^{r( {\rm grav})}_{{\rm ladder}}=\frac{\left(s-\ma^2-\mb^2 \pm \sqrt{(s-\ma^2-\mb^2)^2-4\ma^2\mb^2}\right)^{2(r+1)}}{((s-\ma^2-\mb^2)^2-4\ma^2\mb^2)^{r/2}t}.
\ee

\section{Discussions}

In this work we explored leading singularities (LS) of amplitudes of massive scalar fields interacting via the exchange of gravitons. While leading singularities have been extensively explored in gauge theory and gravity (and their supersymmetric generalizations), the main applications have been for massless external states.

Having massive external particles leads to the presence of classical effects coming from loops in perturbation theory. Classical effects in loop computations are known to originate in certain regions of the loop integration space. Restricting the integration to those regions captures all the classical effects. However, classical contributions can also come accompanied by quantum pieces.

Leading singularities are not directly supported on the regions contributing to classical or quantum pieces. In fact, most leading singularities are computed on contours that are analytic continuations of the loop momenta and do not belong to any physical region. Despite this separation from physical regions, leading singularities capture valuable information about the analytic structure of amplitudes. As explained in sections 2 and 4, leading singularities at one-loop capture information about double discontinuities in the s- and t-channel for the box topology and in the double t-channel for the triangle topology.

In section 3 we found that by using a double dispersion relation construction in the t-channel, the triangle leading singularities were ``stable" under the integration procedure. This means that, up to terms irrelevant to classical scattering, leading singularities preserved their form after the integrals were computed. Using this fact we were able to express the complete one-loop contribution of the amplitude to the classical post-Newtonian expansion purely in terms of leading singularities. We also found that those with a box topology did not contribute to the classical pieces as they did not have double discontinuities in the t-channel.

It is tempting to suggest that this phenomenon can continue at higher loops. More explicitly, it would be interesting to explore the possibility that classical effects are those that are ``stable" under multiple t-channel discontinuities. In fact, even at one-loop, the triangle LS has an infinite number of t-channel discontinuities. In contrast, quantum contributions come from pieces that eventually stop having discontinuities in the t-channel. 

It is well-known that in supersymmetric gauge theories the symbols of the amplitude have been a powerful tool in the study of their analytic structure (for a review see \cite{Golden:2013xva} and \cite{Goncharov:2010jf} for a dramatic simplification obtained by using this technique). The symbol is a mathematical tool designed to store the information of multiple branch cuts and their discontinuities when they are packaged in transcendental functions such as polylogarithms. One could try to formally extend the concept of the symbol to cases of functions with square roots and allow for an infinite number of entries. This would depart from the connection to period integrals but if that were possible, then it is reasonable to expect that classical pieces are those with an infinite-length generalized symbol while quantum ones are those with a finite-length generalized symbol. Even more speculatively, one could even be able to see that functions with only infinite-length or ``classical-like" symbols cannot produce physically meaningful results at short distances and therefore they have to be corrected by finite-length or ``quantum-like" symbols. We leave these intriguing possibilities for future research.

Several directions for future research are clear. The most pressing one is to work out all the leading singularities that contribute at two and three loops to try and reproduce post-Newtonian results. Another direction is the extension to the case of a massless particle of helicity $|h|=0,\frac{1}{2},1$ interacting with a scalar massive one, this case is known as ``light-bending" in the literature and has been addressed in a variety of ways \cite{Bjerrum-Bohr:2014zsa,Bjerrum-Bohr:2014lea,Holstein:2016cfx,Bjerrum-Bohr:2016hpa,Bai:2016ivl}. Also interesting is the extension to massive particles with spin \cite{Porto:2005ac,Holstein:08,Vaidya:2014kza}. Adding spin to the external massive particles usually leads to complications in most approaches but using leading singularities one could expect that they are minimal.

\section*{Acknowledgements}

We would like to thank Matt von Hippel, Barbara Soda, and Jingxiang Wu for collaboration at an early stage of this project which started during the 2017 PSI Winter School. We also thank Sebastian Mizera and Guojun Zhang for useful discussions.  A. G. wishes to express his thanks to Prof. J. Zanelli for support and encouragement, as well as CONICYT for financial support. This research was supported in part by Perimeter Institute for Theoretical Physics. Research at Perimeter Institute is supported by the Government of Canada through the Department of Innovation, Science and Economic Development Canada and by the Province of Ontario through the Ministry of Research, Innovation and Science.

\bibliographystyle{JHEP}
\bibliography{references}

\end{document}